\documentclass[prb,amsmath,amssymb,twocolumn]{revtex4-2}
\usepackage{graphicx}
\usepackage{subfigure}
\usepackage{adjustbox}
\usepackage{bm}
\usepackage{color}
\usepackage{braket}
\usepackage{standalone}
\usepackage{multirow}
\usepackage{tikz}
\usepackage{mathrsfs}
\usepackage{dsfont}
\usepackage{comment}
\usepackage{amsmath}
\usepackage{mathtools}
\usepackage[colorlinks,bookmarks=true,citecolor=blue,linkcolor=red,urlcolor=blue]{hyperref}

\begin{document}
\title{Robust quantum many-body scars in the one-dimensional spin-$1$ Kitaev model}

\author{Sashikanta Mohapatra}
\email{sashikanta@imsc.res.in}
\author{Ajit C. Balram}
\email{cb.ajit@gmail.com}
\affiliation{Institute of Mathematical Sciences, CIT Campus, Chennai 600113, India}
\affiliation{Homi Bhabha National Institute, Training School Complex, Anushaktinagar, Mumbai 400094, India}

\begin{abstract}
Experimental observation of coherent oscillations in a Rydberg atom chain [Bernien \emph{et al.}, Nature {\bf 551}, 579 (2017)] has led to the discovery of quantum many-body scars (QMBS) which is a new paradigm for ergodicity-breaking. The experimental findings in the Rydberg chain can be well captured by a kinetically constrained model called the ``PXP'' model, which has been shown to host the Eigenstate Thermalization Hypothesis (ETH)-violating scar states in the middle of the spectrum. Much effort has been put into identifying similar kinetically restricted systems that show a violation of ETH. In this work, we study the QMBS that can arise in one such model, namely the spin-$1$ Kitaev chain, where owing to some conserved quantities, the Hilbert space gets fragmented into unequal disconnected subspaces. Recently, You \emph{et. al} [Phys. Rev. Research {\bf 4}, 013103 (2022)] showed that the ground state sector of this chain can be mapped exactly onto the prototypical PXP model and thus hosts QMBSs. Here, we demonstrate that the phenomenon of scarring is also present in other sectors, and in particular, we identify a sector that exhibits substantially more scarring than the ground state one. We propose an initial state and numerically demonstrate that its fidelity revivals are robust and longer-lived than those in the PXP model.
\end{abstract}

\maketitle

\section{Introduction}
Rapid improvements in the platforms for realizing and controlling non-equilibrium dynamics of closed quantum systems, such as ultracold atoms~\cite{Kinoshita2006}, trapped ions~\cite{Smith2016}, nitrogen-vacancy centers~\cite{kucsko2018critical}, etc., has enabled a study of the thermalization of quantum systems isolated from external baths. A generic isolated quantum system is expected to be ergodic, i.e., under the unitary dynamics of its Hamiltonian, any initial state would eventually evolve into a featureless thermal state. This loss of information on the initial state's configuration presents a barrier to protecting quantum information. As a result, it is crucial to search for non-ergodic systems that resist thermalization. The Eigenstate Thermalization Hypothesis (ETH)~\cite{deutsch1991quantum, srednicki1994chaos} regulates the characteristics of ergodic quantum systems and describes how far-from-equilibrium initial states evolve in time to reach a final state that is described by a thermal ensemble. ETH suggests \emph{all} the eigenstates of ergodic systems are thermal and thus any initial state evolves into a thermal state at long times. Two well-known exceptions to the ETH paradigm are integrable and many-body localized systems~\cite{BASKO20061126,pal2010many,nandkishore2015many}. In integrable systems, the presence of an extensive number of conserved quantities prevents an initial state from fully exploring all the allowed configurations in the Hilbert space. In MBL systems, the presence of interactions~\cite{Schulz2019stark} and strong disorder~\cite{kjall2014many} leads to an emergent integrability that prevents thermalization. These two ergodicity-breaking mechanisms are of the strong form in that \emph{every} eigenstate exhibits features of an athermal state. 

Recently, experimental findings in an ultracold Rydberg atom chain~\cite{bernien2017probing} revealed a new mechanism for weak ergodicity-breaking. When the Rydberg atoms were initialized in a particular state, the so-called N\'eel state, they do not thermalize and instead display long-lived coherent oscillations. On the other hand, certain other initial states do exhibit thermal behavior. The theoretical description of the Rydberg chain is captured by the kinetically constrained ``PXP" model~\cite{Fendley2004competing}. Since the Rydberg atoms are quite large, it is energetically prohibitive to simultaneously excite two nearest neighboring atoms~\cite{urban2009observation, Lesanovsky2011many}. The `P' in the PXP is a projector that exactly projects out these configurations in which the nearest neighboring sites are both in excited states. This Rydberg blockade constraint imposes a restriction on the allowed configurations for the system which results in a constrained Hilbert space that the system can access. Numerical studies of the PXP model~\cite{Turner2018quantumscar, Turner2018weak,khemani2019signatures, Lin2019exact} have revealed the presence of anomalous states at equidistant energies that have sub-extensive entanglement entropy (EE) in the otherwise thermal bulk spectrum. These special eigenstates obey the area-law of EE rather than the volume-law of EE as anticipated by ETH and have substantial overlap with the N\'eel state which results in the observed coherent revivals. This phenomenon is dubbed quantum many-body scars (QMBS)~\cite{Turner2018quantumscar}.
These scar states are vanishingly rare and typically their number grows only algebraically with system size while the Hilbert space dimension grows exponentially with system size. As a result, these scars only lead to a weak or incomplete breach of ETH.

In recent years, substantial theoretical effort~\cite{moudgalya2021quantum,Choi2019emergent,Chandran2023qmbs,Desaules2021proposal,Khemani2020localization,Wildeboer2022qmbs} has been put in, in tandem with experiments~\cite{Desaules2021proposal, Zhang2023}, to look for systems that can support QMBS. QMBS have been identified in many-body systems, such as the Affleck-Kennedy-Lieb-Tasaki model~\cite{Moudgalya2018exact, Moudgalya2018AKLT, Mark2020AKLT}, integer spin XY model~\cite{Chattopadhyay2020qmbs, Schecter2019weak}, $\eta$-pairing Hubbard model~\cite{Moudgalya2020hubbard, Mark2020hubbard, Bull2019systematic}, thin torus limit of quantum hall phases~\cite{Moudgalya2020thintorus}, tilted 1D Fermi-Hubbard model~\cite{Desaules2021proposal}, etc. In this work, we study QMBS in the spin-$1$ Kitaev chain~\cite{Kitaev2006anyons, Sen2010kitaev}, where previous studies~\cite{You2022qmbs} have demonstrated that the PXP model is embedded in one of its subspaces, thereby making it an ideal candidate system to support QMBS. We study other subspaces (besides the one that has the PXP in it) of this model and see if they too can support QMBS. We identify a sector where the scarring is considerably stronger than that observed in the PXP model. Analogous to the N\'eel state of the PXP model, we propose an initial state in this sector that shows remarkably persistent fidelity oscillations.

The remainder of this paper is organized as follows. We give a brief overview of the one-dimensional (1D) spin-$1$ Kitaev model in Sec.~\ref{sec: 1DKitaev}. In Sec.~\ref{subsec: specialsector} we study a particular sector of this model and its associated constrained dynamics and find that this subspace hosts anomalous scarred states. We identify an initial state in this subspace and numerically demonstrate that it has robust and long-lived coherent oscillations. We show that the forward scattering approximation nicely captures these scarred states. In Sec. \ref{subsec: othersectors} we consider some other subspaces of the Kitaev chain and show that the fidelity oscillations of analogous initial states in these subspaces decay rapidly. Finally, we summarize our results in Sec.~\ref{sec: conclusion} and present an outlook for the future.

\section{The one-dimensional Kitaev model}
\label{sec: 1DKitaev}
The spin-$1$ Kitaev chain can be obtained as a single row of the two-dimensional Kitaev model~\cite{Kitaev2006anyons}. We start with the general spin-$S$ Kitaev model on the honeycomb lattice that is described by the Hamiltonian 
\begin{equation}
\label{Eq: 1}
H_{K}^{\rm 2D}=J_{x} \sum_{{\langle i,j \rangle}_{x}} S_{i}^{x} S_{j}^{x} + J_{y} \sum_{{\langle i,j \rangle}_{y}} S_{i}^{y} S_{j}^{y} + J_{z} \sum_{{\langle i,j \rangle}_{z}} S_{i}^{z} S_{j}^{z},
\end{equation}
where operators $S_{j}^{a}$ (with $a{=}x,y,z$) are the spin-$S$ operators at site $j$ and $\langle i,j \rangle_a$ denotes nearest neighbors in the $a$-direction. The spin operators satisfy the usual $SU(2)$ algebra i.e., $[S_{i}^{a}, S_{j}^{b}]{=}i\delta_{ij}\epsilon_{abc}S_{j}^{c}$, where $\epsilon_{abc}$ is the totally anti-symmetric Levi-Civita tensor. Setting $J_z{=}0$ in Eq.~\eqref{Eq: 1}, we get a set of decoupled 1D chains any one of which of $N$ sites is described by the Hamiltonian~\cite{Sen2010kitaev}
\begin{equation}
\label{Eq: 2}
    H^{\rm 1D}_{K}(\{J\}) =\sum_{j=1}^{N/2}(J_{2j-1} S_{2j-1}^xS_{2j}^x + J_{2j} S_{2j}^y S_{2j+1}^y).
\end{equation}
In general, the coupling constants $J$'s could be different from each other. However, throughout this work, we will consider the simple case where all $J$'s are equal and set to unit strength i.e., $J_{l}=1~\forall l$. Thus, we end up with the following Hamiltonian for the spin-$S$ Kitaev chain
\begin{equation}
\label{Eq: 3}
    H^{\rm 1D}_{K}=\sum_{j=1}^{N/2}(S_{2j-1}^xS_{2j}^x + S_{2j}^y S_{2j+1}^y),
\end{equation}
which is the model that we will work with throughout this paper.
Next, we would like to find the symmetries of the Hamiltonian of Eq.~\eqref{Eq: 3}. To do so, we define site parity operators $\mathcal{P}_{j}^{a}$ on every site as
\begin{equation}
\label{Eq: 4}
    \mathcal{P}^a=e^{i\pi S_j^a}.
\end{equation}
The Ising-like terms in Eq.~\eqref{Eq: 3} change the value of total $S_{z}$ at the sites adjoining a link i.e., $S^{z}_{2j{-}1}{+}S^z_{2j}$ at the $x$-link ($2j{-}1,2j$) and the value of $S^{z}_{2j}{+}S^z_{2j{+}1}$ at the $y$-link ($2j,2j{+}1$), by either $0$ or ${\pm} 2$. Therefore the bond parity operators $\mathcal{B}_{j}$ on odd and even bonds defined by 
\begin{equation}
\label{Eq: 5}
    \mathcal{B}_{2j-1}=\mathcal{P}_{2j-1}^y \mathcal{P}_{2j}^y,\hspace{0.2cm} \text{and} \hspace{0.2cm} \mathcal{B}_{2j}=\mathcal{P}_{2j}^x \mathcal{P}_{2j+1}^x
\end{equation}
remain invariant under the action of Hamiltonian. Thus we have 
\begin{equation}
\label{Eq: 6}
    [\mathcal{B}_j,H]=0,~\forall j,
\end{equation}
and these constitute symmetries of the spin-$S$ Kitaev chain. By performing the following unitary transformation on the even sites~\cite{Sen2010kitaev}
\begin{equation}
\label{Eq: 7}
    S_{2j}^x \rightarrow S_{2j}^y,\hspace{0.5cm} S_{2j}^y \rightarrow S_{2j}^x \hspace{0.5cm} S_{2j}^z \rightarrow -S_{2j}^z,
\end{equation}
the Hamiltonian can be cast into the following convenient translationally invariant form
\begin{equation}
    H=\sum_{j=1}^{N} S_j^xS_{j+1}^y.
    \label{Eq: 8}
\end{equation}
Upon the unitary transformation of Eq.~\eqref{Eq: 7}, the bond parity operators take the universal form (independent of whether bond $j$ is even or odd)
\begin{equation}
\label{Eq: 9}
    \mathcal{B}_j=\mathcal{P}_j^y \mathcal{P}_{j+1}^x.
\end{equation}

From here on, we shall restrict ourselves to the spin-$1$ case of our interest and work with its natural representation given by the orthonormal basis states $\{|x\rangle,|y\rangle,|z\rangle\}$ defined as
\begin{equation}
    \begin{split}
        \ket{x}\equiv\frac{1}{\sqrt{2}}(\ket{-1}-\ket{1}), \ket{y}\equiv\frac{i}{\sqrt{2}}(\ket{-1}+\ket{1}), \ket{z }\equiv\ket{0},
    \end{split}
    \label{Eq:10}
\end{equation}
where $|m\rangle$ is the eigenstate of the spin-$1$ operator $S_{i}^{z}$ with eigenvalue $m{=}{-}1,0,1$. In this representation the spin-$1$ operators can be written as $S_{bc}^a{=} i\epsilon_{abc}$ and their matrix representation is
\begin{equation}
    \begin{split}
        S^x= 
        \begin{pmatrix}
            0 & 0 & 0\\
            0& 0 & -i\\
            0& i & 0
        \end{pmatrix}, 
        S^y= 
        \begin{pmatrix}
            0 & 0 & i\\
            0& 0 & 0\\
            -i& 0 & 0
        \end{pmatrix}, 
        S^z= 
        \begin{pmatrix}
            0 & -i & 0\\
            i& 0 & 0\\
            0& 0 & 0
        \end{pmatrix}.
    \end{split}
    \label{Eq: 11}
\end{equation}
Furthermore, the $3\times 3$ matrices corresponding to the site parity operators $\mathcal{P}^{a}$ of Eq.~\eqref{Eq: 4} are diagonal and given by
\begin{equation}
    \begin{split}
        \mathcal{P}^x{=}
        \begin{pmatrix}
            1 & 0 & 0\\
            0 & {-}1 & 0\\
            0 & 0 & {-}1
        \end{pmatrix}, 
        \mathcal{P}^y{=}
        \begin{pmatrix}
            {-}1 & 0 & 0\\
            0 & 1 & 0\\
            0 & 0 & {-}1
        \end{pmatrix}, 
        \mathcal{P}^z{=}
        \begin{pmatrix}
            {-}1 & 0 & 0\\
            0 & {-}1 & 0\\
            0 & 0 & 1
        \end{pmatrix}.
    \end{split}
    \label{Eq: 12}
\end{equation}
From this matrix representation, we can readily read off that the eigenvalues of the operators $\mathcal{P}_{j}^{a}$ are ${\pm}1$ with the eigenvalue ${-}1$ being doubly degenerate. Therefore, the eigenvalues of bond parity operators $\mathcal{B}_{j}$ defined in Eq.~\eqref{Eq: 9} are also $b_{j}{=}{\pm} 1$ since they are just products of the site parity operators. Moreover, as the site-parity operators $\mathcal{P}_{j}^{a}$ are diagonal, they commute with each other. The bond operators $\mathcal{B}_{j}$ being products of diagonal site-parity operators are also diagonal and commute with each other [along with the fact that they commute with the Hamiltonian as shown in Eq.~\eqref{Eq: 6}]. This implies the Hilbert space can be decomposed into $2^{N}$ sectors (of unequal sizes since the eigenvalue ${-}1$ is doubly degenerate) and each sector can be represented by a set of bond invariants $\vec{b}{\equiv}\{b_{1},b_{2},{\cdots},b_{N}\}$.

Projection into these sectors imposes restrictions on the allowed configurations of two neighboring sites. For the nearest neighbor sites $\langle j,j{+}1 \rangle$ there are a total of $3{\times}3{=}9$ allowed states which, based on the eigenvalue of the bond operator $\mathcal{B}_{j}$, get fragmented into the following two sets
\begin{equation}    \ket{xy},\hspace{0.2cm}\ket{xz},\hspace{0.2cm}\ket{yx},\hspace{0.2cm}\ket{zy}\hspace{0.2cm}\text{and}\hspace{0.2cm}\ket{zz} \hspace{0.2cm}\text{have}\hspace{0.2cm} b_j{=}1, 
    \label{Eq: 13}
\end{equation}
and
\begin{equation}
  |{xx}\rangle,\hspace{0.2cm}|{yy}\rangle,\hspace{0.2cm}|{yz}\rangle\hspace{0.2cm}\text{and}\hspace{0.2cm}|{zx}\rangle\hspace{0.2cm}\text{have}\hspace{0.2cm} b_j{=}{-}1.
    \label{Eq: 14}
\end{equation}
The existence of these constrained subspaces makes the spin-$1$ Kitaev chain a viable candidate to host QMBS.

\section{QMBS in the spin-$1$ Kitaev chain}
\label{sec: qmbs_subspaces}
\begin{figure}[h]
    \centering
    \includegraphics[scale=0.31]{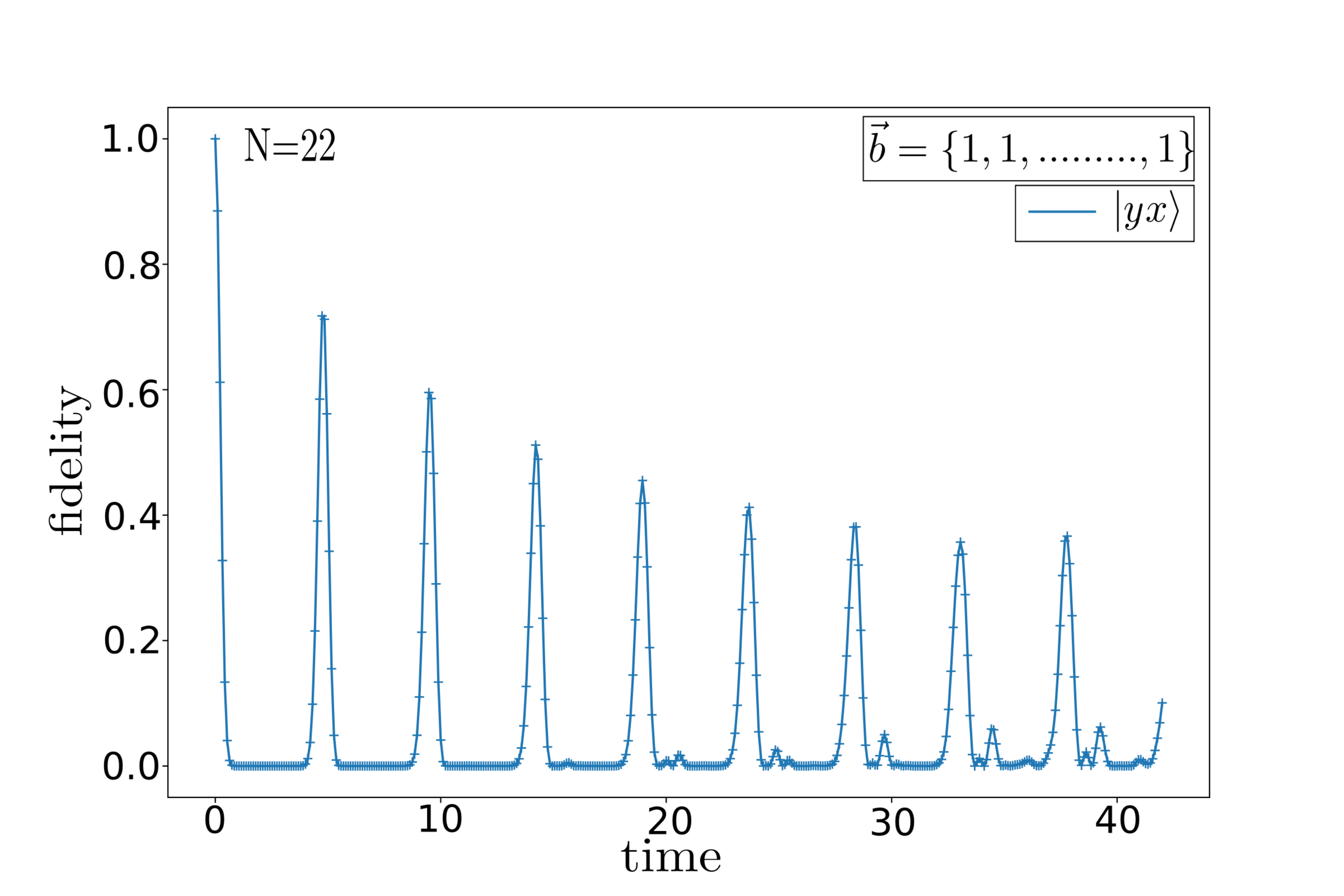}
    \caption{Fidelity of the $|Z_2\rangle_{\rm Kitaev}$ state showing periodic revivals. Data are for $N{=}22$ sites with periodic boundary conditions. The dimension of the sector $D(1,1,\cdots,1){=}39,603$. }
    \label{fig: 1}
\end{figure}

The authors of Ref.~\cite{Sen2010kitaev} showed that the ground state of the Hamiltonian of Eq.~\eqref{Eq: 8} lies in the subspace with all $b_{j}{=}1$. The restriction on the neighboring sites in this sector exactly mimics the Rydberg blockade constraint~\cite{Sen2010kitaev, You2022qmbs}. 
Thus, the $\vec{b}{=}\{1,1,\cdots,1\}$ subspace can exactly be mapped into the PXP model (see App.~\ref{sec: app_mapping_PXP_spin_1_Kitaev}) and therefore hosts QMBS~\cite{You2022qmbs}. The corresponding N\'eel state for the spin-$1$ chain is given by $|Z_{2}\rangle_{\rm Kitaev}{=}|yx\rangle{\equiv}|yxyx\cdots yx\rangle$ and the fidelity for this state $F(t){=}|\langle{Z_2}|\exp(-iHt)|{Z_2}\rangle|^{2}$ gives rise to coherent oscillation as shown in Fig.~\ref{fig: 1}.

We will show in the subsequent sections that some other subspaces of the spin-$1$ Kitaev chain also harbor scarred eigenstates, though in general, it is difficult to find the corresponding spin-$1/2$ Hamiltonian like the PXP one as it involves complicated forms with long-range interactions. In particular, we find that the $\vec{b}{=}\{1,1,{-}1,1,1,{-}1,\cdots,1,1,{-}1\}$ sector exhibits a more pronounced scarring effect than the ground state one and we will discuss the fate of QMBS in this sector next.

\subsection{The $\vec{b}{=}\{1,1,{-}1,1,1-1,{\cdots},1,1,{-}1\}$ sector}
\label{subsec: specialsector}
\begin{figure}
     \centering
     \includegraphics[scale=0.3]{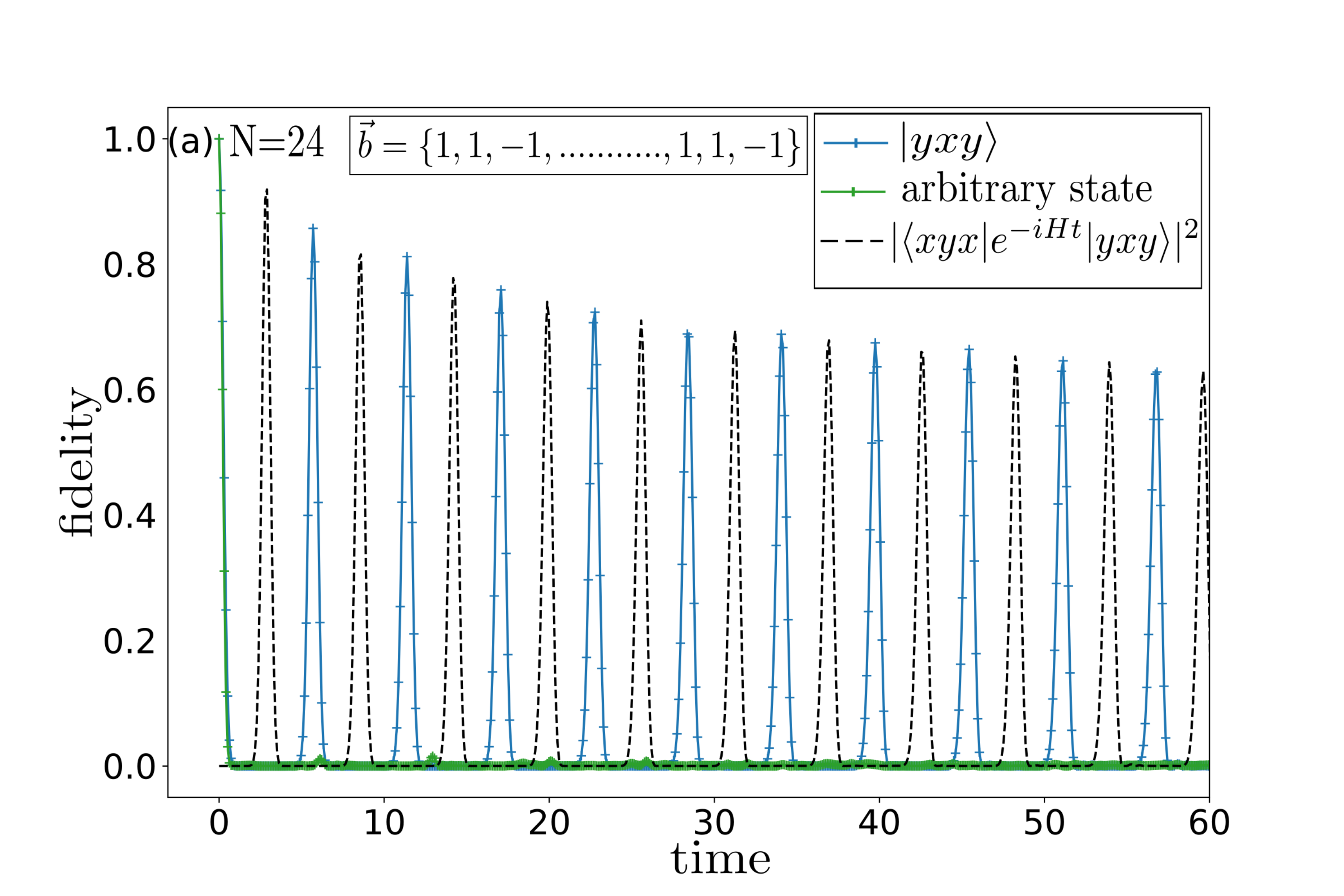} \\
         \includegraphics[scale=0.3]{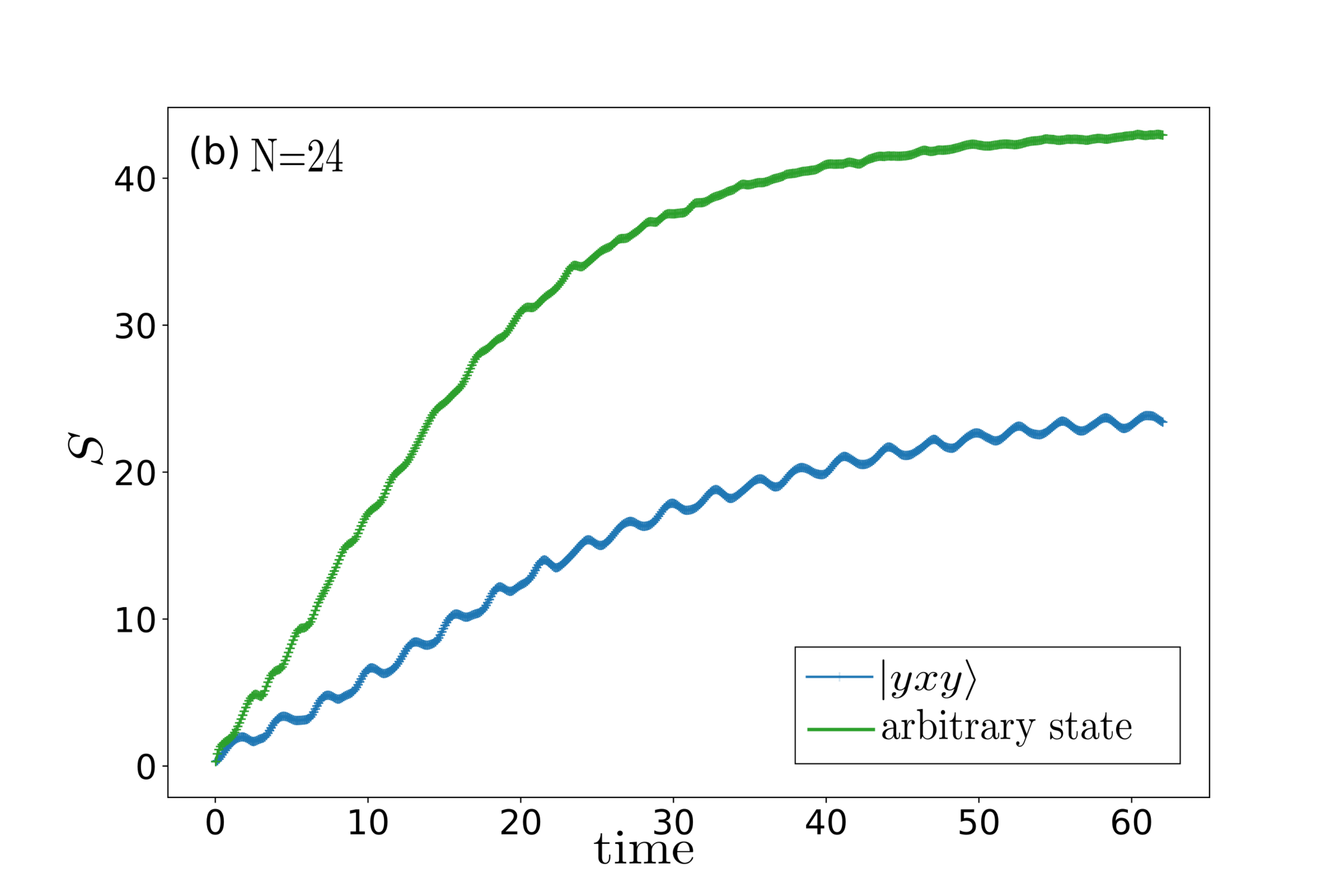}
        \caption{(a) Fidelity dynamics for the initial state $|yxy\rangle$ and a randomly chosen product state. The former shows coherent oscillations whereas the latter thermalizes rapidly. The black dotted lines show the probability of state transfer between $|yxy\rangle$ and $|xyx\rangle$ (b) Entanglement entropy S for the same initial states. The Entropy of the $|yxy\rangle$ state grows linearly with time but it also shows oscillations due to scarring. Data is shown for N=24 sites with PBC.}
        \label{fig: 3}
\end{figure}

We first unravel the structure of the constrained Hilbert space of this sector. There are two types of constraints on the states of nearest neighboring sites: i) since $b_{3j}{=}{-}1$, there are four possible states given in Eq.~\eqref{Eq: 14} that the neighboring sites $\langle3j,3j{+}1\rangle$ can be in, whereas, ii) since $b_{k}{=}1~\forall k{\neq 3j}$, there are five possible states given in Eq.~\eqref{Eq: 13} that the neighboring sites $\langle k,k{+}1\rangle$, $k{\neq 3j}$ can be in. The dimension of Hilbert space $\mathcal{H}$ of this sector is known to be $D(1,1,{-}1,1,1,{-}1,{\cdots},1,1,{-}1){\approx}1.55113^{N}$ for a system of size $N$ with periodic boundary condition (PBC)~\cite{Sen2010kitaev}. In this sector, the Hamiltonian transforms the state $|{yz}\rangle$ to $|{zx}\rangle$ and vice-versa over the bond with $b_{j}{=}{-}1$ and $|{zz}\rangle$ to $ |{yx}\rangle$ and vice-versa over the bond with $b_j{=}1$. Therefore the Hilbert space of this sector can be constructed by taking any initial state of the sector as root state (call it $|{R}\rangle)$ and successively applying the Hamiltonian on it, i.e., 
\begin{equation}
    \mathcal{H}_{\{1,1,-1,1,1-1,\cdots,1,1-1\}} \equiv \text{Span}\{|{R}\rangle,H|{R}\rangle,H^2 |{R}\rangle,\cdots\}.
    \label{Eq. 15}
\end{equation}
\begin{figure}[h]
    \centering
    \includegraphics[scale=0.3]{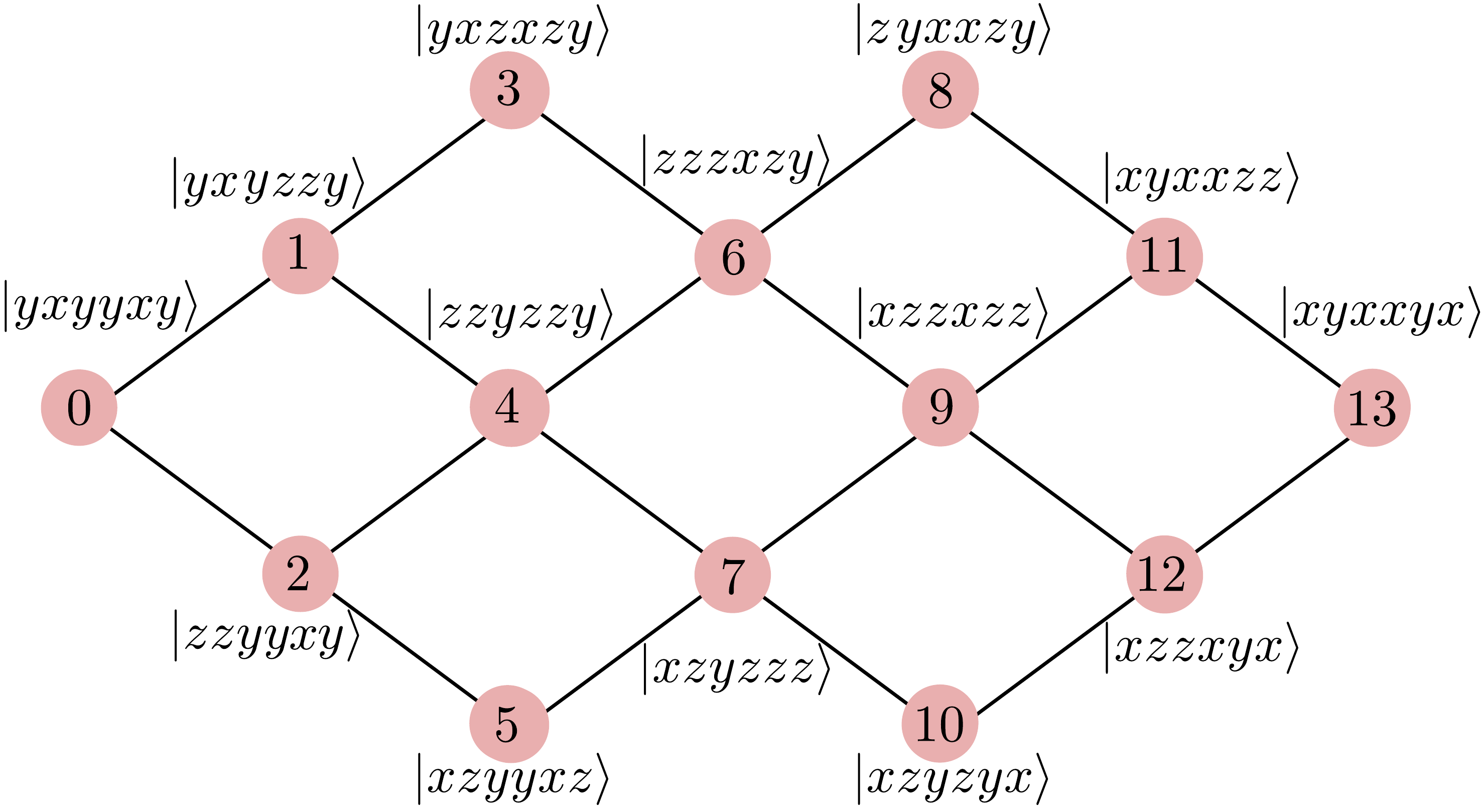}
    \caption{The action of the Kitaev Hamiltonian of Eq.~\eqref{Eq: 8} on the Hilbert space of the sector $\vec{b}{=}\{1,1,-1,1,1,-1\}$ for $N{=}6$ sites with periodic boundary conditions. }
    \label{fig: 2}
\end{figure}
Fig.~\ref{fig: 2} shows the constrained Hilbert space and the action of the Hamiltonian in this sector for $N{=}6$ sites with PBC. In the graph, each node corresponds to a product state of the subspace and the edges connect the configurations that result from a given product state due to the action of the Hamiltonian. This graph representation will be helpful in the forward scattering approximation (FSA) defined later in this section.
We now study the dynamics of the basis states of this subspace using the exact diagonalization of the Hamiltonian. We find that initial states of the kind $|yxy\rangle{\equiv}|yxyyxy\cdots yxy\rangle$ and $|xyx\rangle{\equiv}|xyxxyx\cdots xyx\rangle$ show long-lived revivals. Fig.~\ref{fig: 3}(a) depicts the evolution of the initial state $|yxy\rangle$ and a randomly chosen product state under the Kitaev Hamiltonian with $N{=}24$ sites. The $|yxy\rangle$ state shows the hallmark of QMBS wherein fidelity oscillations are robust and long-lived. In particular, the fidelity oscillations for this state are more robust (peaks heights are higher as evidenced by the fact that the first revival peak displays ${>}80\%$ return probability to the initial state) and longer-lived (persist for a longer time) as compared to that of the $|Z_2\rangle_\text{Kitaev}$ state shown in Fig.~\ref{fig: 1}. In sharp contrast, a random state thermalizes rapidly [see the green curve shown in Fig.~\ref{fig: 3}(a)].

We can visualize the scarred dynamics in this sector as the state bouncing between the two corner states $|yxy\rangle$ and $|xyx\rangle$ of the Hilbert space graph shown in Fig. \ref{fig: 2}. The dotted line in Fig.~\ref{fig: 3}(a), where we plot $|\langle xyx|e^{-iHt}|yxy \rangle|^2$, which is the probability of finding the state in $|xyx\rangle$ following time evolution from the initial state $|yxy\rangle$, illustrates this back-and-forth motion. In Fig.~\ref{fig: 3}(b) we plot the growth of EE with time for different initial states. The EE of a subregion $A$ is defined as the von Neumann entropy of the reduced density matrix of the subsystem as $S_{A}{=}{-}\text{Tr$_{A^{c}}$}\{\rho_{A}\text{ln}\rho_{A}\}$, where $\rho_{A}$ is the reduced density matrix of subsystem $A$ and the trace is taken over its complement $A^{c}$. For the $|yxy\rangle$ state, along with an increase as a function of time, the EE mirrors the oscillations seen in the fidelity. Moreover, the rate at which EE grows in the $|yxy\rangle$ state is much smaller as compared to that of a randomly chosen state, suggesting that the initial $|yxy\rangle$ state results in non-ergodic behavior.

\begin{figure}[h]
    \centering
    \includegraphics[scale=0.3]{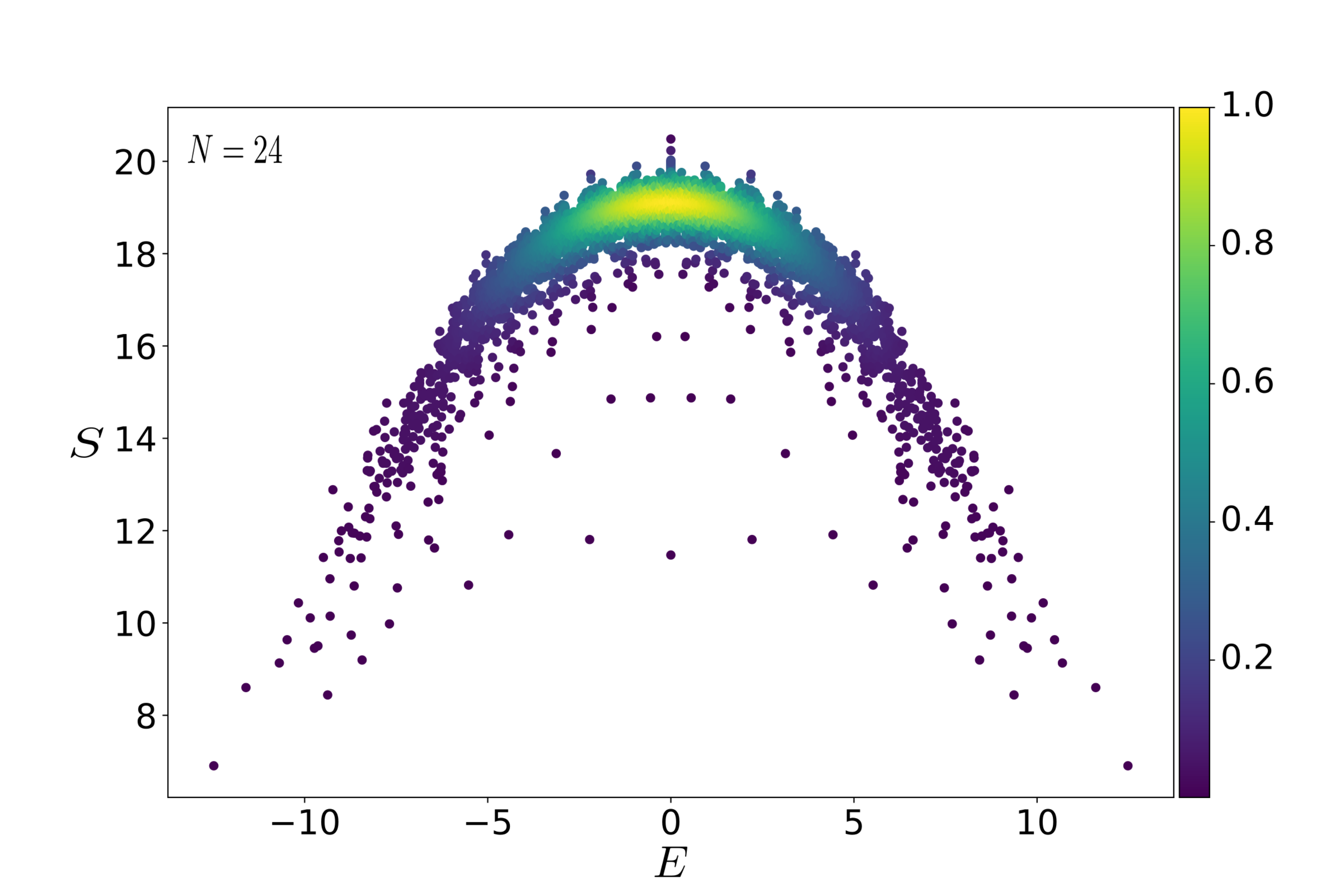}
    \caption{Bipartite (equipartition) entanglement entropy of the eigenstates of the $\vec{b}{=}\{1,1,{-}1,1,1-1,{\cdots},1,1,{-}1\}$ in the spin-$1$ Kitaev model as a function of energy. The bulk states satisfy the volume law of EE, however, there are several states that reside over the entire range of the spectrum and carry low EE and thereby violate ETH. Data is shown for $N{=}24$ sites with PBC. The color scale on the right indicates the density of the data points.}
    \label{fig: 4}
\end{figure}

Thermalization and its breakdown can also be probed by measuring the spread of EE of eigenstates. ETH predicts a ``volume-law" scaling of EE, i.e., for a 1D system EE scales linearly with the size of the subsystem. Fig.~\ref{fig: 4} demonstrates that the bipartite (equipartitioned) $S$ for the majority of the eigenstates of the $\vec{b}{=}\{1,1,{-}1,1,1-1,{\cdots},1,1,{-}1\}$ sector do exhibit the volume-law behavior that is consistent with the prediction of ETH. However, in addition to the bulk of highly entangled states, there are outliers over the entire range of the spectrum that have much lower entropy that violates the volume law predicted by ETH. 

\begin{figure}
    \centering
    \includegraphics[scale=0.105]{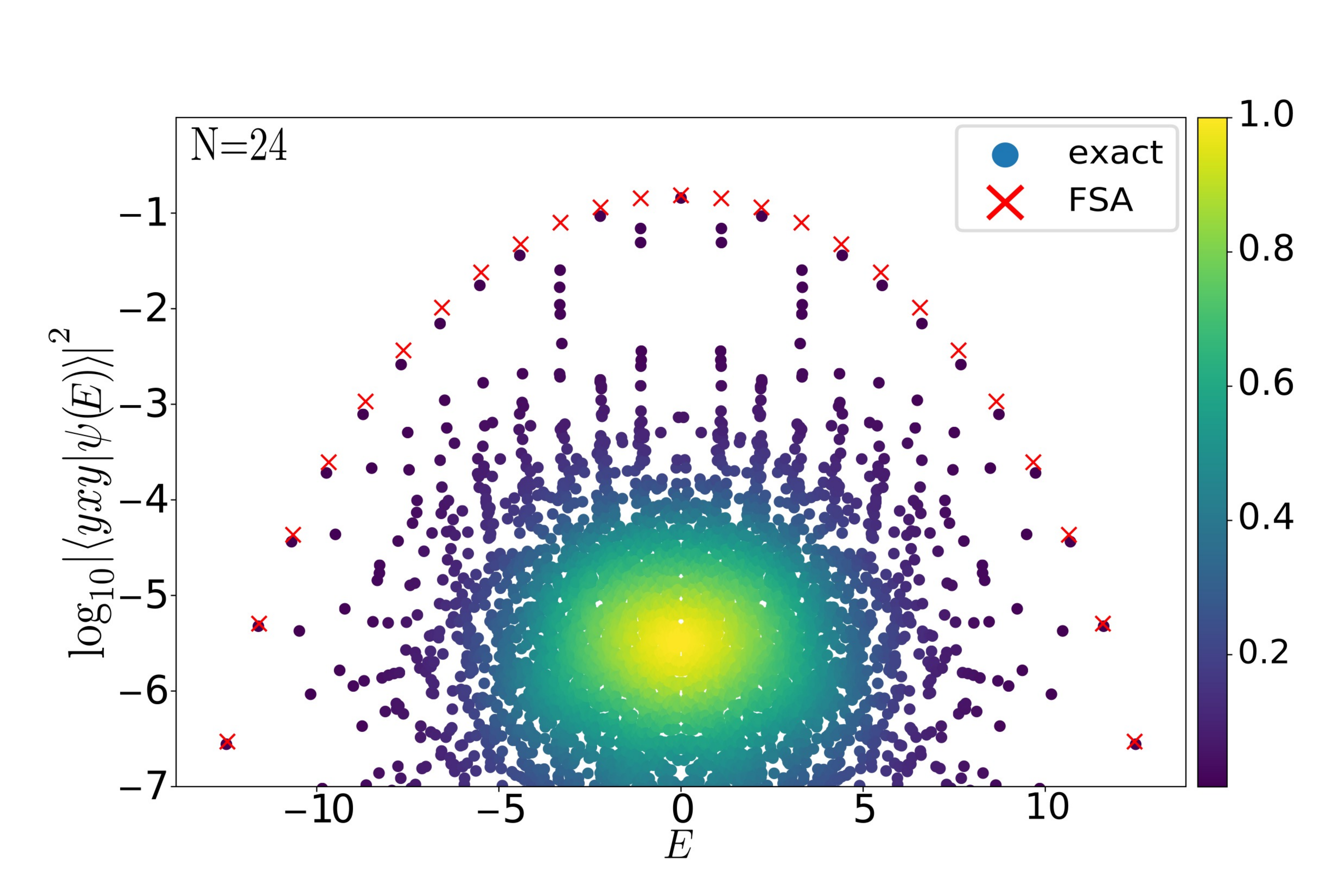}
    \caption{Density plot showing the overlap of the initial product state $|{yxy}\rangle$ with the eigenstates of the $\vec{b}{=}\{1,1,{-}1,1,1-1,{\cdots},1,1,{-}1\}$ sector in the spin-$1$ Kitaev model. The cross marks denote the overlap with the eigenstates of the FSA Hamiltonian [see Eq.~\eqref{Eq: 20}], which very well approximate the topmost states in the towers of scar states.}
    \label{fig: 5}
\end{figure}

The fidelity oscillations observed in Fig.~\ref{fig: 3}(a) arise precisely due to the existence of these relatively small number of athermal eigenstates that are spread throughout the bulk spectrum but carry low EE. These scarred states have anomalously high overlap with the initial product state $|{yxy}\rangle$ as shown in Fig.~\ref{fig: 5}. Like in other models hosting QMBS~\cite{Turner2018quantumscar, Moudgalya2020thintorus}, the projection onto the $|{yxy}\rangle$ state displays towers of special \emph{equispaced} (with the spacing in energy determining the inverse time period of oscillations seen in the fidelity) eigenstates having anomalously high overlap with the initial product state $|{yxy}\rangle$. The observed coherent oscillations in fidelity, sub-thermal entanglement entropy of certain eigenstates, and anomalously large overlap of these eigenstates with a particular initial product state results in the non-ergodic dynamics and establishes the existence of QMBS in this subspace of spin-$1$ Kitaev chain.

Furthermore, as has been demonstrated for the PXP mode, the topmost state in the towers of scarred eigenstates in this sector can be well-approximated using the so-called Forward Scattering Approximation (FSA)~\cite{Turner2018quantumscar}. The FSA mechanism involves constructing an approximate Hamiltonian whose eigenstates reproduce the scarred states. We start by splitting the Hamiltonian into forward and backward propagating parts as $H{=}H^{{+}} {+} H^{{-}}$, where
\begin{equation}
    \begin{split}
         H^+{=}\sum_{i=1,4,7,\cdots} |{yx}\rangle\langle{zz}| + \sum_{i=2,5,6,\cdots} |{zz}\rangle\langle{yx}|\\ - \sum_{i=1,4,7,\cdots} |{yz}\rangle\langle{zx}|
    \end{split}
    \label{Eq: 16}
\end{equation}
\begin{equation}
    \begin{split}
         H^-{=}\sum_{i=1,4,7,\cdots} |{zz}\rangle\langle{yx}| + \sum_{i=2,5,6,\cdots} |{yx}\rangle\langle{zz}|\\ - \sum_{i=1,4,7,\cdots} |{zx}\rangle\langle{yz}|.
    \end{split}
    \label{Eq: 17}
\end{equation}

Then we construct the basis vectors $|{0}\rangle,|{1}\rangle,\cdots,|{N}\rangle$ of the effective Hamiltonian $H_{\rm FSA}$, where $|{0}\rangle{\equiv}|{yxy}\rangle$ and $|{n}\rangle{=}(1/\sqrt{c_{n}})(H^{+})^{n} |{yxy}\rangle$ ($c_{n}$ is the normalization constant). In the Hilbert space graph shown in Fig.~\ref{fig: 2}, the action of $H^{+}$ corresponds to moving from left to right (right to left for $H^{-}$) and $H^{+}$ annihilates the $|xyx\rangle$ state ($H^{-}$ annihilates the $|yxy\rangle$ state). Therefore, starting from the $|yxy\rangle$ state the FSA recurrence closes after $N{+}1$ steps once the forward propagation reaches the $|xyx\rangle$ state at the opposite end of the Hilbert space graph shown in Fig.~\ref{fig: 2}. The approximation in FSA entails that the Hilbert space of basis states $|0\rangle,|1\rangle,\cdots,|N\rangle$ is closed under the action of the Kitaev Hamiltonian of Eq. \eqref{Eq: 8}. The action of $H$ on these basis states is given by
 \begin{equation}
     \begin{split}
         H|n\rangle=&H^+|n\rangle+H^- |n\rangle\\
         =& \beta_{n+1} |n+1\rangle + H^- |n\rangle,
     \end{split}
     \label{Eq: 18}
 \end{equation}
where $\beta_{n}{=}\sqrt{c_{n}/c_{n-1}}$. Thus to make the Hilbert space closed under the action of the Hamiltonian we have to approximate 
\begin{equation}
    H^-|n\rangle \approx \beta_n |n-1\rangle.
    \label{Eq: 19}
\end{equation}
Using Eqs.~\eqref{Eq: 18} and \eqref{Eq: 19} the Hamiltonian takes the form of the following tridiagonal matrix which is the FSA Hamiltonian
\begin{equation}
    H_{\rm FSA}=
    \begin{pmatrix}
        0   & \beta_1 &   &   &  \\
        \beta_1 & 0  & \beta_2 &    &   \\
            & \beta_2 & 0  & \ddots &  \\
            &   & \ddots &\ddots & \beta_N \\
            &   &   &\beta_N & 0
    \end{pmatrix}.
    \label{Eq: 20}
\end{equation}
As shown by the cross marks in Fig.~\ref{fig: 5}, the eigenstates of the FSA Hamiltonian of Eq.~\eqref{Eq: 20} provide an excellent approximation to the special scarred eigenstates of the Kitaev Hamiltonian in the sector $\vec{b}{=}\{1,1,{-}1,1,1,{-}1,{\cdots},1,1,{-}1\}$.

\subsection{QMBS in other sectors}
\label{subsec: othersectors}
\begin{figure}[h]
    \includegraphics[scale=0.3]{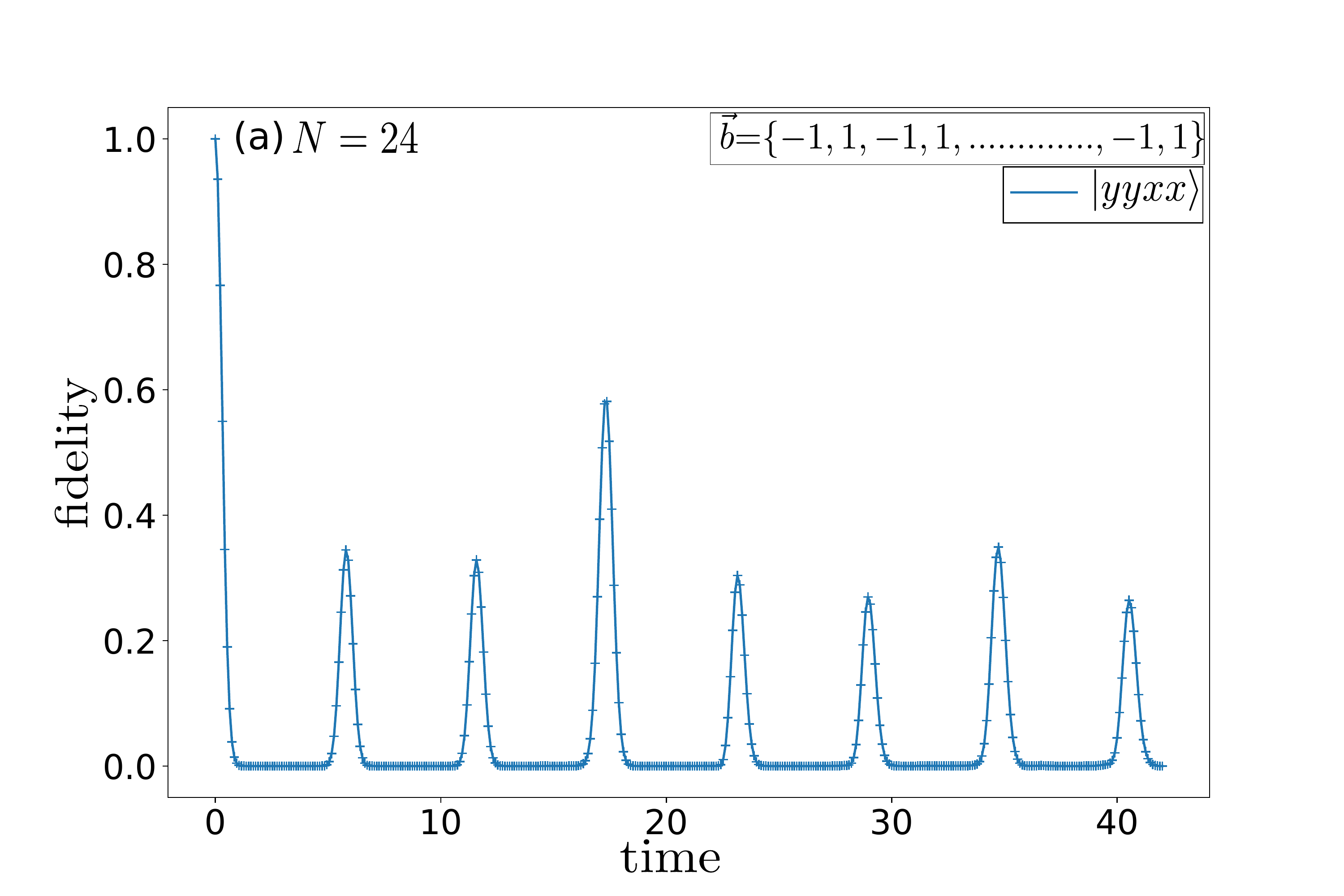}\\
    \includegraphics[scale=0.3]{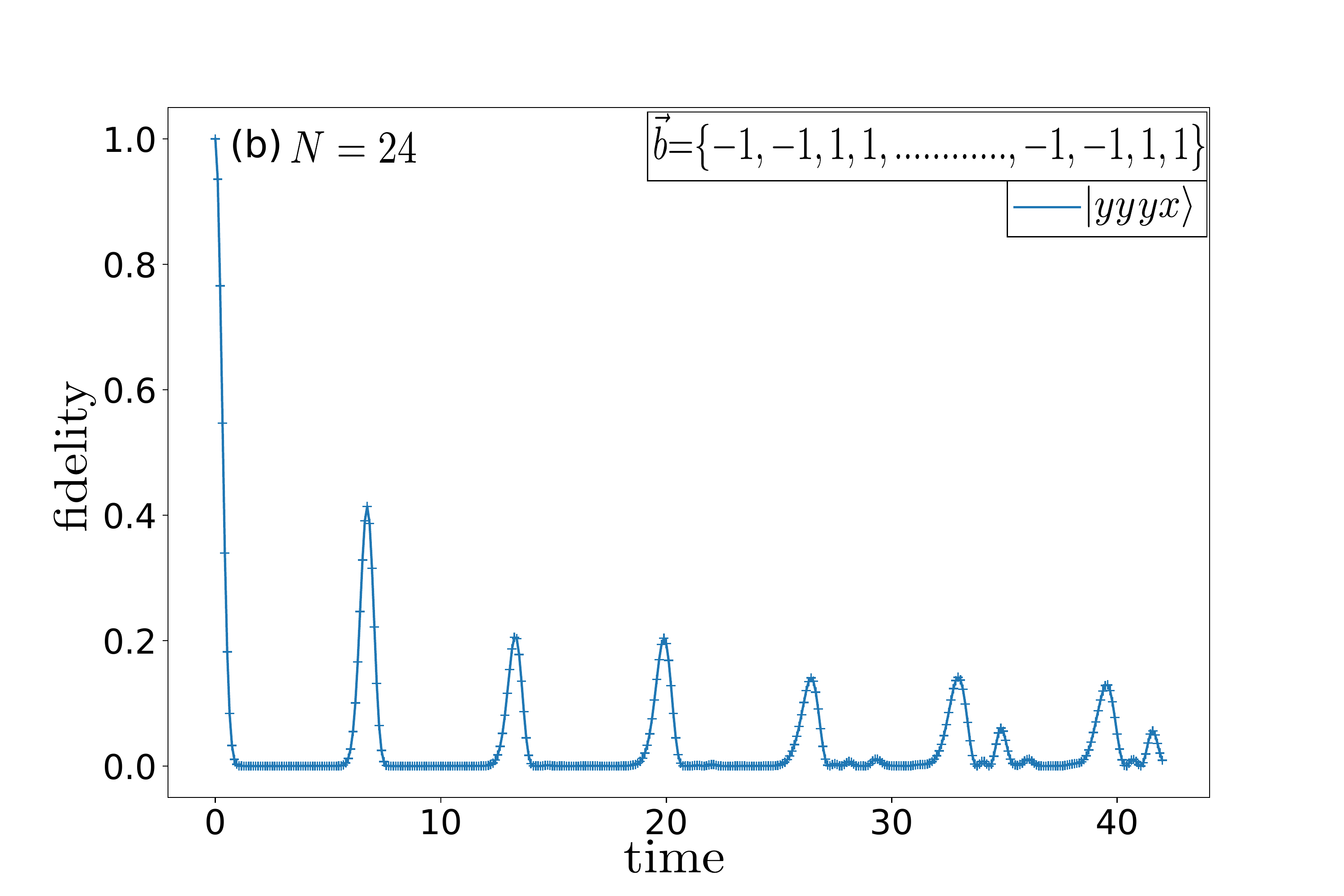}
    \caption{Fidelity of the state (a) $|yyxx\rangle$ which lies in the sector $\vec{b}{=}\{{-}1,1,{-}1,1,{\cdots},{-}1,1\}$ (b) $|yyyx\rangle$ which lies in the sector $\vec{b}{=}\{{-}1,{-}1,1,1,{-}1,{-}1,1,1,{\cdots},{-}1,{-}1,1,1\}$. The fidelity oscillations are weaker (peak heights are reduced) and decay faster in these sectors as compared to the ones shown in Fig.~\ref{fig: 3}. Data are shown for $N{=}24$ sites. }
    \label{fig: 6}
\end{figure}

\begin{figure}
    \includegraphics[scale=0.29]{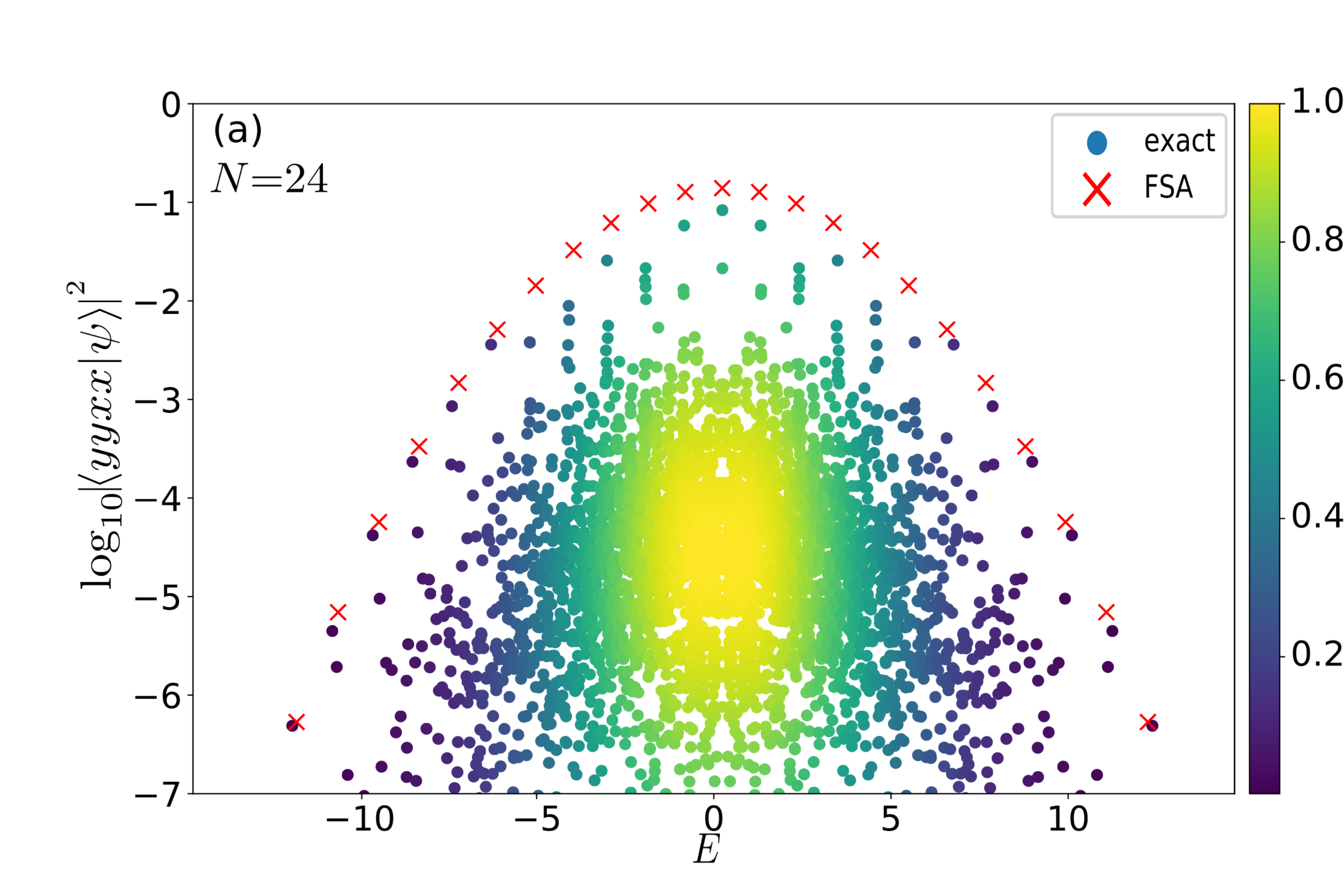}\\
    \includegraphics[scale=0.29]{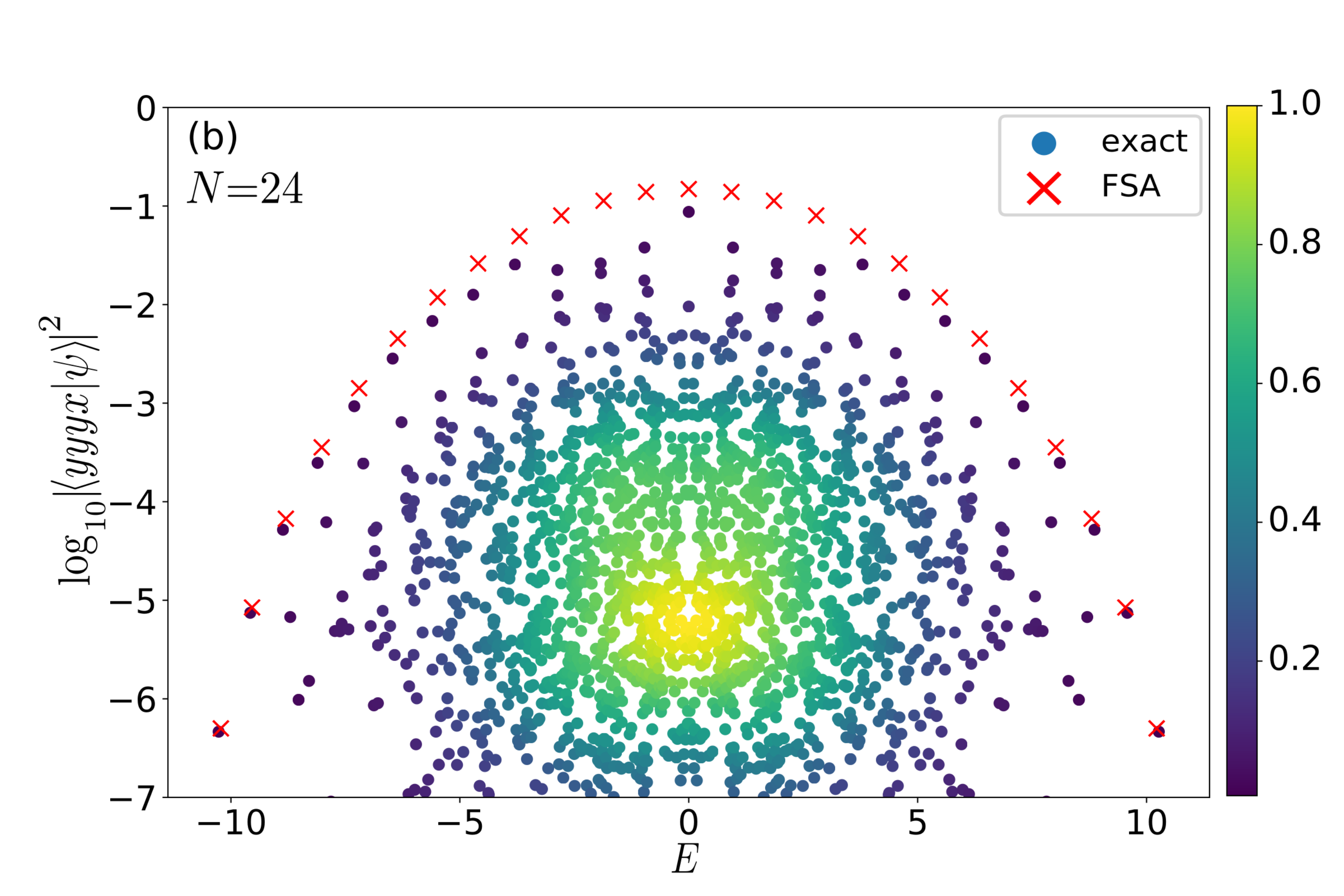}
    \caption{Density plot of the overlap of the initial product states that show fidelity oscillations in particular sectors (see Fig.~\ref{fig: 6}) of the spin-$1$ Kitaev model with all eigenstates in that sector. The crosses show the overlap with eigenstates of the forward-scattering approximation Hamiltonian $H_{\rm FSA}$. The overlap of the scar states with the product states fails to match the magnitude as predicted by FSA which is consistent with the observation of a faster decay of fidelity in these states. Data are shown for (a) $\vec{b}{=}\{{-}1,1,{-}1,1,\cdots,{-}1,1\}$  and (b) $\vec{b}{=}\{{-}1,{-}1,1,1,{-}1,{-}1,1,1,\cdots,{-}1,{-}1,1,1\}$ sectors of the spin-$1$ Kitaev chain of $N{=}24$ sites. }
    \label{fig: 7}
\end{figure}

We have also looked for the possibility of scarring in other sectors of the Kitaev chain by studying the dynamics from different initial product states. Amongst all the initial states and sectors we considered, we found that initial product states $|yyxx\rangle{\equiv}|yyxxyyxx\cdots yyxx\rangle$ in the sector $\vec{b}{=}\{{-}1,1,{-}1,1,\cdots,{-}1,1\}$  and the state $|yyyx\rangle{\equiv}|yyyxyyyx\cdots yyyx\rangle$ in the sector $\vec{b}{=}\{{-}1,{-}1,1,1,{-}1,{-}1,1,1,\cdots,{-}1,{-}1,1,1\}$ also show oscillations in the fidelity. In Fig.~\ref{fig: 6} we plot the return probabilities for the aforementioned initial states. We note that the oscillations are weaker (as evidenced by the peak heights) and decay much faster in these sectors. One way to understand this is that the FSA does not work well in these sectors as shown in Fig.~\ref{fig: 7}.

The authors of Ref.~\cite{Choi2019emergent} showed that Hamiltonians hosting QMBS support an emergent approximate $SU(2)$ symmetry within a subspace of the Hilbert space. The revivals from the initial product states can then be thought of as the coherent rotation of the large $SU(2)$ degree of freedom. In the $SU(2)$ algebra, $H^{+}$ and $H^{-}$ act as an analog of the raising and lowering operators. Their commutator $H^{z}{=}[H^{+}, H^{-}]$ acts as $S^{z}$ operator, and the FSA states are its eigenstates. However, in Eq.~\eqref{Eq: 19} we saw that $H^{-}$ only approximately inverts the action of $H^{+}$, which is why the algebra is not exact and perfect revivals are not observed. We expect that if the FSA gives a good representation of the exact scar states, then one sees strong revivals. Otherwise, if the FSA does not represent the scar states well, the fidelity decays quickly. This is consistent with our numerical observations. 

These results also show that it is not necessarily the case that the more constrained a Hilbert space is the more scarring it shows. In general, the more the number of $b_j{=}{-}1$ larger the number of constraints and the fewer the number of states in the corresponding Hilbert space. Nevertheless, as we have shown above, the more constrained sector $\vec{b}{=}\{{-}1,{-}1,1,1,{-}1,{-}1,1,1,{\cdots}\}$ shows less scarring than the less constrained $\vec{b}{=}\{1,1,{-}1,1,1,{-}1,{\cdots}\}$ sector. The strength of scarring is determined by the structure of the graph of the corresponding Hilbert space and how well the FSA works there. We note here that we have checked that all sectors in the spin-$1$ Kitaev chain that we considered do not show Poison level statistics which rules out an integrability-based explanation for the athermal behavior we observe.

\section{Summary and Conclusion}
\label{sec: conclusion}
In this paper, we studied the time evolution of initial states in certain sectors of the spin-$1$ Kitaev chain, where the Hilbert space is fragmented into $2^{N}$ unequal subspaces. We looked at the dynamics of the initial states in these constrained subspaces. We found that the $|yxy\rangle$ state in the $\vec{b}{=}\{1,1,{-}1,1,1,{-}1,\cdots\}$ sector showed the most prominent coherent oscillations in fidelity when evolved under the Kitaev Hamiltonian, more so than even the $|Z_{2}\rangle_{\rm Kitaev}$ state of the celebrated PXP model which is embedded in the $\vec{b}{=}\{1,1,\cdots,1\}$ sector that hosts the ground state of the spin-$1$ Kitaev chain. The coherent dynamics in the $\vec{b}{=}\{1,1,{-}1,1,1,{-}1,\cdots\}$ sector were characterized by special eigenstates that have anomalously low entanglement entropy, and high overlap with the initial $|yxy\rangle$ state. We also showed that these special scarred states can be well-approximated by the FSA. Furthermore using the FSA, we showed why certain other sectors do not show long-lived oscillations in fidelity. It would be interesting to see if the scarring phenomena we observed can be understood using analytical techniques such as the recently proposed broken unitary picture of dynamics in QMBS~\cite{Rozon22_broken_unitary}, or interpreting it as a one-dimensional chiral scattering problem~\cite{Windt22_SqueezingQMBS}, or projector-embedding~\cite{Shiraishi17_projector_embedding} or commutant algebras~\cite{Moudgalya22_commutant_algebras}.

The Kitaev chain provides a model system and framework to study the dynamics of constrained systems. Here, we only looked at the spin-$1$ chain and it would be interesting to look at higher spins and see whether they exhibit QMBS. Another potential avenue that could be worth studying in the future is to explore the existence of QMBS in higher dimensions and/or in different geometries.

\section*{Acknowledgments}
We acknowledge useful discussions with Diptiman Sen, Kartiek Agarwal, Sanjay Moudgalya, and Zlatko Papi\'c. Computational portions of this research work were conducted using the Nandadevi supercomputer, which is maintained and supported by the Institute of Mathematical Science’s High-Performance Computing Center.

\appendix 
\section{Mapping the $\vec{b}{=}\{1,1,{\cdots},1\}$ sector to the PXP model}
\label{sec: app_mapping_PXP_spin_1_Kitaev}
In this appendix, we show that the $\vec{b}{=}\{1,1,{\cdots},1\}$ sector of the spin-$1$ Kitaev chain can be mapped to the PXP model. Since all $b_{j}{=}1$ in this sector, there are five allowed states for any pair of nearest neighbor sites $\langle j,j{+}1\rangle$ as shown in Eq.~\eqref{Eq: 13}. Owing to these constraints, configurations of neighboring sites can be written in terms of spin-$1/2$ degrees of freedom on the dual lattice (for each bond between sites $j$ and $j{+}1$ on the primal lattice, on the dual lattice we define a site at $\{j{+}1/2\}$) of $N$ sites using the following mapping
\begin{equation}
    \begin{split}
    \label{Eq: A1}
        &|yx\rangle_{j,j+1} \rightarrow |\uparrow\rangle_{j+\frac{1}{2}}\\
        &|zy\rangle_{j,j+1} \rightarrow |\downarrow\rangle_{j+\frac{1}{2}}\\
        &|xz\rangle_{j,j+1} \rightarrow |\downarrow\rangle_{j+\frac{1}{2}}\\
        &|zz\rangle_{j,j+1} \rightarrow |\downarrow\rangle_{j+\frac{1}{2}}\\
        &|xy\rangle_{j,j+1} \rightarrow |\downarrow\rangle_{j+\frac{1}{2}}.
    \end{split}
\end{equation}
This mapping does not allow the nearest neighbors on the dual lattice to be in the configuration $|{\uparrow}{\uparrow}\rangle$ which is precisely the Rydberg blockade constraint (no two nearest neighbors are simultaneously in the excited state) that is implemented in the PXP model. Though the mapping in Eq.~\eqref{Eq: A1} appears to be many-to-one it is not. The reverse mapping from the dual lattice to the spin-$1$ primal lattice is given by
\begin{equation}
    \begin{split}
        &|\downarrow\uparrow \rangle_{j-\frac{1}{2},j+\frac{1}{2}}\rightarrow|y\rangle_{j} \\
         &|\uparrow\downarrow \rangle_{j-\frac{1}{2},j+\frac{1}{2}}\rightarrow|x\rangle_{j} \\
          &|\downarrow\downarrow \rangle_{j-\frac{1}{2},j+\frac{1}{2}}\rightarrow|z\rangle_{j},
    \end{split}
    \label{Eq: A2}
\end{equation}
which ensures that the mapping is one-to-one. Note that a similar mapping was used in Ref.~\cite{Moudgalya2020thintorus} to map the thin-torus limit of the pair-hoping Hamiltonian of the $\nu{=}1/3$ fractional quantum hall effect to the PXP model. With this mapping, the (non-vanishing) action of the spin-$1$ Kitaev model on the primal lattice leads to the following terms in the Hamiltonian $\mathcal{H}_{\{1,1,{\cdots},1\}}$ in the dual space
\begin{equation}
    \begin{split}
        H_{j,j+1}|\ast \overset{j}{y}\overset{j+1}{x} \ast \rangle = | \ast \overset{j}{z}\overset{j+1}{z} \ast \rangle\\
        \implies \mathcal{H}_{j+\frac{1}{2}} |\downarrow \overset{j+\frac{1}{2}}{\uparrow} \downarrow\rangle = |\downarrow\overset{j+\frac{1}{2}}{\downarrow}\downarrow\rangle\\
        H_{j,j+1}|\ast \overset{j}{z}\overset{j+1}{z} \ast \rangle = | \ast \overset{j}{y}\overset{j+1}{x} \ast \rangle\\
        \implies \mathcal{H}_{j+\frac{1}{2}} |\downarrow \overset{j+\frac{1}{2}}{\downarrow} \downarrow\rangle = |\downarrow\overset{j+\frac{1}{2}}{\uparrow}\downarrow\rangle,
    \end{split}
    \label{Eq: A3}
\end{equation}
where $|\ast\rangle$ corresponds to any allowed configuration on the sites that respect the above-mentioned constraints. The terms in the Hamiltonian of Eq.~\eqref{Eq: A3} are exactly the ones that appear in the PXP model, which is given by
\begin{equation}
    \mathcal{H}_{\{1,1,\cdots,1\}}=\sum_{j=1}^N P_{j-1}\sigma_j^x P_{j+1},
    \label{Eq: A4}
\end{equation}
where the Pauli operators are defined in the usual way with $\sigma^{x}{=}(|{\downarrow}\rangle\langle{\uparrow}|{+}|{\uparrow}\rangle\langle{\downarrow}|)$ and $\sigma^{z}{=}(|{\uparrow}\rangle\langle{\uparrow}|{-}|{\downarrow}\rangle\langle{\downarrow}|)$, and the projectors $P_{j}{=}(1{-}\sigma_j^{z})/2$ ensure that no two spin-$1/2$ nearest neighbors are simultaneously in the excited $|{\uparrow}\rangle$ state. 
Thus, the spin-$1$ Kitaev chain Hamiltonian restricted to the $\vec{b}{=}\{1,1,\cdots,1\}$ sector (which hosts its ground state) exactly maps into the PXP model.

\bibliography{references}

\begin{thebibliography}{41}%
\makeatletter
\providecommand \@ifxundefined [1]{%
 \@ifx{#1\undefined}
}%
\providecommand \@ifnum [1]{%
 \ifnum #1\expandafter \@firstoftwo
 \else \expandafter \@secondoftwo
 \fi
}%
\providecommand \@ifx [1]{%
 \ifx #1\expandafter \@firstoftwo
 \else \expandafter \@secondoftwo
 \fi
}%
\providecommand \natexlab [1]{#1}%
\providecommand \enquote  [1]{``#1''}%
\providecommand \bibnamefont  [1]{#1}%
\providecommand \bibfnamefont [1]{#1}%
\providecommand \citenamefont [1]{#1}%
\providecommand \href@noop [0]{\@secondoftwo}%
\providecommand \href [0]{\begingroup \@sanitize@url \@href}%
\providecommand \@href[1]{\@@startlink{#1}\@@href}%
\providecommand \@@href[1]{\endgroup#1\@@endlink}%
\providecommand \@sanitize@url [0]{\catcode `\\12\catcode `\$12\catcode
  `\&12\catcode `\#12\catcode `\^12\catcode `\_12\catcode `\%12\relax}%
\providecommand \@@startlink[1]{}%
\providecommand \@@endlink[0]{}%
\providecommand \url  [0]{\begingroup\@sanitize@url \@url }%
\providecommand \@url [1]{\endgroup\@href {#1}{\urlprefix }}%
\providecommand \urlprefix  [0]{URL }%
\providecommand \Eprint [0]{\href }%
\providecommand \doibase [0]{https://doi.org/}%
\providecommand \selectlanguage [0]{\@gobble}%
\providecommand \bibinfo  [0]{\@secondoftwo}%
\providecommand \bibfield  [0]{\@secondoftwo}%
\providecommand \translation [1]{[#1]}%
\providecommand \BibitemOpen [0]{}%
\providecommand \bibitemStop [0]{}%
\providecommand \bibitemNoStop [0]{.\EOS\space}%
\providecommand \EOS [0]{\spacefactor3000\relax}%
\providecommand \BibitemShut  [1]{\csname bibitem#1\endcsname}%
\let\auto@bib@innerbib\@empty
\bibitem [{\citenamefont {Kinoshita}\ \emph {et~al.}(2006)\citenamefont
  {Kinoshita}, \citenamefont {Wenger},\ and\ \citenamefont
  {Weiss}}]{Kinoshita2006}%
  \BibitemOpen
  \bibfield  {author} {\bibinfo {author} {\bibfnamefont {T.}~\bibnamefont
  {Kinoshita}}, \bibinfo {author} {\bibfnamefont {T.}~\bibnamefont {Wenger}},\
  and\ \bibinfo {author} {\bibfnamefont {D.~S.}\ \bibnamefont {Weiss}},\ }\href
  {https://doi.org/10.1038/nature04693} {\bibfield  {journal} {\bibinfo
  {journal} {Nature}\ }\textbf {\bibinfo {volume} {440}},\ \bibinfo {pages}
  {900} (\bibinfo {year} {2006})}\BibitemShut {NoStop}%
\bibitem [{\citenamefont {Smith}\ \emph {et~al.}(2016)\citenamefont {Smith},
  \citenamefont {Lee}, \citenamefont {Richerme}, \citenamefont {Neyenhuis},
  \citenamefont {Hess}, \citenamefont {Hauke}, \citenamefont {Heyl},
  \citenamefont {Huse},\ and\ \citenamefont {Monroe}}]{Smith2016}%
  \BibitemOpen
  \bibfield  {author} {\bibinfo {author} {\bibfnamefont {J.}~\bibnamefont
  {Smith}}, \bibinfo {author} {\bibfnamefont {A.}~\bibnamefont {Lee}}, \bibinfo
  {author} {\bibfnamefont {P.}~\bibnamefont {Richerme}}, \bibinfo {author}
  {\bibfnamefont {B.}~\bibnamefont {Neyenhuis}}, \bibinfo {author}
  {\bibfnamefont {P.~W.}\ \bibnamefont {Hess}}, \bibinfo {author}
  {\bibfnamefont {P.}~\bibnamefont {Hauke}}, \bibinfo {author} {\bibfnamefont
  {M.}~\bibnamefont {Heyl}}, \bibinfo {author} {\bibfnamefont {D.~A.}\
  \bibnamefont {Huse}},\ and\ \bibinfo {author} {\bibfnamefont
  {C.}~\bibnamefont {Monroe}},\ }\href {https://doi.org/10.1038/nphys3783}
  {\bibfield  {journal} {\bibinfo  {journal} {Nature Physics}\ }\textbf
  {\bibinfo {volume} {12}},\ \bibinfo {pages} {907} (\bibinfo {year}
  {2016})}\BibitemShut {NoStop}%
\bibitem [{\citenamefont {Kucsko}\ \emph {et~al.}(2018)\citenamefont {Kucsko},
  \citenamefont {Choi}, \citenamefont {Choi}, \citenamefont {Maurer},
  \citenamefont {Zhou}, \citenamefont {Landig}, \citenamefont {Sumiya},
  \citenamefont {Onoda}, \citenamefont {Isoya}, \citenamefont {Jelezko},
  \citenamefont {Demler}, \citenamefont {Yao},\ and\ \citenamefont
  {Lukin}}]{kucsko2018critical}%
  \BibitemOpen
  \bibfield  {author} {\bibinfo {author} {\bibfnamefont {G.}~\bibnamefont
  {Kucsko}}, \bibinfo {author} {\bibfnamefont {S.}~\bibnamefont {Choi}},
  \bibinfo {author} {\bibfnamefont {J.}~\bibnamefont {Choi}}, \bibinfo {author}
  {\bibfnamefont {P.~C.}\ \bibnamefont {Maurer}}, \bibinfo {author}
  {\bibfnamefont {H.}~\bibnamefont {Zhou}}, \bibinfo {author} {\bibfnamefont
  {R.}~\bibnamefont {Landig}}, \bibinfo {author} {\bibfnamefont
  {H.}~\bibnamefont {Sumiya}}, \bibinfo {author} {\bibfnamefont
  {S.}~\bibnamefont {Onoda}}, \bibinfo {author} {\bibfnamefont
  {J.}~\bibnamefont {Isoya}}, \bibinfo {author} {\bibfnamefont
  {F.}~\bibnamefont {Jelezko}}, \bibinfo {author} {\bibfnamefont
  {E.}~\bibnamefont {Demler}}, \bibinfo {author} {\bibfnamefont {N.~Y.}\
  \bibnamefont {Yao}},\ and\ \bibinfo {author} {\bibfnamefont {M.~D.}\
  \bibnamefont {Lukin}},\ }\href
  {https://doi.org/10.1103/PhysRevLett.121.023601} {\bibfield  {journal}
  {\bibinfo  {journal} {Phys. Rev. Lett.}\ }\textbf {\bibinfo {volume} {121}},\
  \bibinfo {pages} {023601} (\bibinfo {year} {2018})}\BibitemShut {NoStop}%
\bibitem [{\citenamefont {Deutsch}(1991)}]{deutsch1991quantum}%
  \BibitemOpen
  \bibfield  {author} {\bibinfo {author} {\bibfnamefont {J.~M.}\ \bibnamefont
  {Deutsch}},\ }\href {https://doi.org/10.1103/PhysRevA.43.2046} {\bibfield
  {journal} {\bibinfo  {journal} {Phys. Rev. A}\ }\textbf {\bibinfo {volume}
  {43}},\ \bibinfo {pages} {2046} (\bibinfo {year} {1991})}\BibitemShut
  {NoStop}%
\bibitem [{\citenamefont {Srednicki}(1994)}]{srednicki1994chaos}%
  \BibitemOpen
  \bibfield  {author} {\bibinfo {author} {\bibfnamefont {M.}~\bibnamefont
  {Srednicki}},\ }\href {https://doi.org/10.1103/PhysRevE.50.888} {\bibfield
  {journal} {\bibinfo  {journal} {Physical review e}\ }\textbf {\bibinfo
  {volume} {50}},\ \bibinfo {pages} {888} (\bibinfo {year} {1994})}\BibitemShut
  {NoStop}%
\bibitem [{\citenamefont {Basko}\ \emph {et~al.}(2006)\citenamefont {Basko},
  \citenamefont {Aleiner},\ and\ \citenamefont {Altshuler}}]{BASKO20061126}%
  \BibitemOpen
  \bibfield  {author} {\bibinfo {author} {\bibfnamefont {D.}~\bibnamefont
  {Basko}}, \bibinfo {author} {\bibfnamefont {I.}~\bibnamefont {Aleiner}},\
  and\ \bibinfo {author} {\bibfnamefont {B.}~\bibnamefont {Altshuler}},\ }\href
  {https://doi.org/https://doi.org/10.1016/j.aop.2005.11.014} {\bibfield
  {journal} {\bibinfo  {journal} {Annals of Physics}\ }\textbf {\bibinfo
  {volume} {321}},\ \bibinfo {pages} {1126} (\bibinfo {year}
  {2006})}\BibitemShut {NoStop}%
\bibitem [{\citenamefont {Pal}\ and\ \citenamefont {Huse}(2010)}]{pal2010many}%
  \BibitemOpen
  \bibfield  {author} {\bibinfo {author} {\bibfnamefont {A.}~\bibnamefont
  {Pal}}\ and\ \bibinfo {author} {\bibfnamefont {D.~A.}\ \bibnamefont {Huse}},\
  }\href {https://doi.org/10.1103/PhysRevB.82.174411} {\bibfield  {journal}
  {\bibinfo  {journal} {Phys. Rev. B}\ }\textbf {\bibinfo {volume} {82}},\
  \bibinfo {pages} {174411} (\bibinfo {year} {2010})}\BibitemShut {NoStop}%
\bibitem [{\citenamefont {Nandkishore}\ and\ \citenamefont
  {Huse}(2015)}]{nandkishore2015many}%
  \BibitemOpen
  \bibfield  {author} {\bibinfo {author} {\bibfnamefont {R.}~\bibnamefont
  {Nandkishore}}\ and\ \bibinfo {author} {\bibfnamefont {D.~A.}\ \bibnamefont
  {Huse}},\ }\href {https://doi.org/10.1146/annurev-conmatphys-031214-014726}
  {\bibfield  {journal} {\bibinfo  {journal} {Annual Review of Condensed Matter
  Physics}\ }\textbf {\bibinfo {volume} {6}},\ \bibinfo {pages} {15} (\bibinfo
  {year} {2015})},\ \Eprint
  {https://arxiv.org/abs/https://doi.org/10.1146/annurev-conmatphys-031214-014726}
  {https://doi.org/10.1146/annurev-conmatphys-031214-014726} \BibitemShut
  {NoStop}%
\bibitem [{\citenamefont {Schulz}\ \emph {et~al.}(2019)\citenamefont {Schulz},
  \citenamefont {Hooley}, \citenamefont {Moessner},\ and\ \citenamefont
  {Pollmann}}]{Schulz2019stark}%
  \BibitemOpen
  \bibfield  {author} {\bibinfo {author} {\bibfnamefont {M.}~\bibnamefont
  {Schulz}}, \bibinfo {author} {\bibfnamefont {C.~A.}\ \bibnamefont {Hooley}},
  \bibinfo {author} {\bibfnamefont {R.}~\bibnamefont {Moessner}},\ and\
  \bibinfo {author} {\bibfnamefont {F.}~\bibnamefont {Pollmann}},\ }\href
  {https://doi.org/10.1103/PhysRevLett.122.040606} {\bibfield  {journal}
  {\bibinfo  {journal} {Phys. Rev. Lett.}\ }\textbf {\bibinfo {volume} {122}},\
  \bibinfo {pages} {040606} (\bibinfo {year} {2019})}\BibitemShut {NoStop}%
\bibitem [{\citenamefont {Kj\"all}\ \emph {et~al.}(2014)\citenamefont
  {Kj\"all}, \citenamefont {Bardarson},\ and\ \citenamefont
  {Pollmann}}]{kjall2014many}%
  \BibitemOpen
  \bibfield  {author} {\bibinfo {author} {\bibfnamefont {J.~A.}\ \bibnamefont
  {Kj\"all}}, \bibinfo {author} {\bibfnamefont {J.~H.}\ \bibnamefont
  {Bardarson}},\ and\ \bibinfo {author} {\bibfnamefont {F.}~\bibnamefont
  {Pollmann}},\ }\href {https://doi.org/10.1103/PhysRevLett.113.107204}
  {\bibfield  {journal} {\bibinfo  {journal} {Phys. Rev. Lett.}\ }\textbf
  {\bibinfo {volume} {113}},\ \bibinfo {pages} {107204} (\bibinfo {year}
  {2014})}\BibitemShut {NoStop}%
\bibitem [{\citenamefont {Bernien}\ \emph {et~al.}(2017)\citenamefont
  {Bernien}, \citenamefont {Schwartz}, \citenamefont {Keesling}, \citenamefont
  {Levine}, \citenamefont {Omran}, \citenamefont {Pichler}, \citenamefont
  {Choi}, \citenamefont {Zibrov}, \citenamefont {Endres}, \citenamefont
  {Greiner}, \citenamefont {Vuleti{\'{c}}},\ and\ \citenamefont
  {Lukin}}]{bernien2017probing}%
  \BibitemOpen
  \bibfield  {author} {\bibinfo {author} {\bibfnamefont {H.}~\bibnamefont
  {Bernien}}, \bibinfo {author} {\bibfnamefont {S.}~\bibnamefont {Schwartz}},
  \bibinfo {author} {\bibfnamefont {A.}~\bibnamefont {Keesling}}, \bibinfo
  {author} {\bibfnamefont {H.}~\bibnamefont {Levine}}, \bibinfo {author}
  {\bibfnamefont {A.}~\bibnamefont {Omran}}, \bibinfo {author} {\bibfnamefont
  {H.}~\bibnamefont {Pichler}}, \bibinfo {author} {\bibfnamefont
  {S.}~\bibnamefont {Choi}}, \bibinfo {author} {\bibfnamefont {A.~S.}\
  \bibnamefont {Zibrov}}, \bibinfo {author} {\bibfnamefont {M.}~\bibnamefont
  {Endres}}, \bibinfo {author} {\bibfnamefont {M.}~\bibnamefont {Greiner}},
  \bibinfo {author} {\bibfnamefont {V.}~\bibnamefont {Vuleti{\'{c}}}},\ and\
  \bibinfo {author} {\bibfnamefont {M.~D.}\ \bibnamefont {Lukin}},\ }\href
  {https://doi.org/10.1038/nature24622} {\bibfield  {journal} {\bibinfo
  {journal} {Nature}\ }\textbf {\bibinfo {volume} {551}},\ \bibinfo {pages}
  {579} (\bibinfo {year} {2017})}\BibitemShut {NoStop}%
\bibitem [{\citenamefont {Fendley}\ \emph {et~al.}(2004)\citenamefont
  {Fendley}, \citenamefont {Sengupta},\ and\ \citenamefont
  {Sachdev}}]{Fendley2004competing}%
  \BibitemOpen
  \bibfield  {author} {\bibinfo {author} {\bibfnamefont {P.}~\bibnamefont
  {Fendley}}, \bibinfo {author} {\bibfnamefont {K.}~\bibnamefont {Sengupta}},\
  and\ \bibinfo {author} {\bibfnamefont {S.}~\bibnamefont {Sachdev}},\ }\href
  {https://doi.org/10.1103/PhysRevB.69.075106} {\bibfield  {journal} {\bibinfo
  {journal} {Phys. Rev. B}\ }\textbf {\bibinfo {volume} {69}},\ \bibinfo
  {pages} {075106} (\bibinfo {year} {2004})}\BibitemShut {NoStop}%
\bibitem [{\citenamefont {Urban}\ \emph {et~al.}(2009)\citenamefont {Urban},
  \citenamefont {Johnson}, \citenamefont {Henage}, \citenamefont {Isenhower},
  \citenamefont {Yavuz}, \citenamefont {Walker},\ and\ \citenamefont
  {Saffman}}]{urban2009observation}%
  \BibitemOpen
  \bibfield  {author} {\bibinfo {author} {\bibfnamefont {E.}~\bibnamefont
  {Urban}}, \bibinfo {author} {\bibfnamefont {T.~A.}\ \bibnamefont {Johnson}},
  \bibinfo {author} {\bibfnamefont {T.}~\bibnamefont {Henage}}, \bibinfo
  {author} {\bibfnamefont {L.}~\bibnamefont {Isenhower}}, \bibinfo {author}
  {\bibfnamefont {D.~D.}\ \bibnamefont {Yavuz}}, \bibinfo {author}
  {\bibfnamefont {T.~G.}\ \bibnamefont {Walker}},\ and\ \bibinfo {author}
  {\bibfnamefont {M.}~\bibnamefont {Saffman}},\ }\href
  {https://doi.org/10.1038/nphys1178} {\bibfield  {journal} {\bibinfo
  {journal} {Nature Physics}\ }\textbf {\bibinfo {volume} {5}},\ \bibinfo
  {pages} {110} (\bibinfo {year} {2009})}\BibitemShut {NoStop}%
\bibitem [{\citenamefont {Lesanovsky}(2011)}]{Lesanovsky2011many}%
  \BibitemOpen
  \bibfield  {author} {\bibinfo {author} {\bibfnamefont {I.}~\bibnamefont
  {Lesanovsky}},\ }\href {https://doi.org/10.1103/PhysRevLett.106.025301}
  {\bibfield  {journal} {\bibinfo  {journal} {Phys. Rev. Lett.}\ }\textbf
  {\bibinfo {volume} {106}},\ \bibinfo {pages} {025301} (\bibinfo {year}
  {2011})}\BibitemShut {NoStop}%
\bibitem [{\citenamefont {Turner}\ \emph
  {et~al.}(2018{\natexlab{a}})\citenamefont {Turner}, \citenamefont
  {Michailidis}, \citenamefont {Abanin}, \citenamefont {Serbyn},\ and\
  \citenamefont {Papi\ifmmode~\acute{c}\else
  \'{c}\fi{}}}]{Turner2018quantumscar}%
  \BibitemOpen
  \bibfield  {author} {\bibinfo {author} {\bibfnamefont {C.~J.}\ \bibnamefont
  {Turner}}, \bibinfo {author} {\bibfnamefont {A.~A.}\ \bibnamefont
  {Michailidis}}, \bibinfo {author} {\bibfnamefont {D.~A.}\ \bibnamefont
  {Abanin}}, \bibinfo {author} {\bibfnamefont {M.}~\bibnamefont {Serbyn}},\
  and\ \bibinfo {author} {\bibfnamefont {Z.}~\bibnamefont
  {Papi\ifmmode~\acute{c}\else \'{c}\fi{}}},\ }\href
  {https://doi.org/10.1103/PhysRevB.98.155134} {\bibfield  {journal} {\bibinfo
  {journal} {Phys. Rev. B}\ }\textbf {\bibinfo {volume} {98}},\ \bibinfo
  {pages} {155134} (\bibinfo {year} {2018}{\natexlab{a}})}\BibitemShut
  {NoStop}%
\bibitem [{\citenamefont {Turner}\ \emph
  {et~al.}(2018{\natexlab{b}})\citenamefont {Turner}, \citenamefont
  {Michailidis}, \citenamefont {Abanin}, \citenamefont {Serbyn},\ and\
  \citenamefont {Papi{\'{c}}}}]{Turner2018weak}%
  \BibitemOpen
  \bibfield  {author} {\bibinfo {author} {\bibfnamefont {C.~J.}\ \bibnamefont
  {Turner}}, \bibinfo {author} {\bibfnamefont {A.~A.}\ \bibnamefont
  {Michailidis}}, \bibinfo {author} {\bibfnamefont {D.~A.}\ \bibnamefont
  {Abanin}}, \bibinfo {author} {\bibfnamefont {M.}~\bibnamefont {Serbyn}},\
  and\ \bibinfo {author} {\bibfnamefont {Z.}~\bibnamefont {Papi{\'{c}}}},\
  }\href {https://doi.org/10.1038/s41567-018-0137-5} {\bibfield  {journal}
  {\bibinfo  {journal} {Nature Physics}\ }\textbf {\bibinfo {volume} {14}},\
  \bibinfo {pages} {745} (\bibinfo {year} {2018}{\natexlab{b}})}\BibitemShut
  {NoStop}%
\bibitem [{\citenamefont {Khemani}\ \emph {et~al.}(2019)\citenamefont
  {Khemani}, \citenamefont {Laumann},\ and\ \citenamefont
  {Chandran}}]{khemani2019signatures}%
  \BibitemOpen
  \bibfield  {author} {\bibinfo {author} {\bibfnamefont {V.}~\bibnamefont
  {Khemani}}, \bibinfo {author} {\bibfnamefont {C.~R.}\ \bibnamefont
  {Laumann}},\ and\ \bibinfo {author} {\bibfnamefont {A.}~\bibnamefont
  {Chandran}},\ }\href {https://doi.org/10.1103/PhysRevB.99.161101} {\bibfield
  {journal} {\bibinfo  {journal} {Phys. Rev. B}\ }\textbf {\bibinfo {volume}
  {99}},\ \bibinfo {pages} {161101} (\bibinfo {year} {2019})}\BibitemShut
  {NoStop}%
\bibitem [{\citenamefont {Lin}\ and\ \citenamefont
  {Motrunich}(2019)}]{Lin2019exact}%
  \BibitemOpen
  \bibfield  {author} {\bibinfo {author} {\bibfnamefont {C.-J.}\ \bibnamefont
  {Lin}}\ and\ \bibinfo {author} {\bibfnamefont {O.~I.}\ \bibnamefont
  {Motrunich}},\ }\href {https://doi.org/10.1103/PhysRevLett.122.173401}
  {\bibfield  {journal} {\bibinfo  {journal} {Phys. Rev. Lett.}\ }\textbf
  {\bibinfo {volume} {122}},\ \bibinfo {pages} {173401} (\bibinfo {year}
  {2019})}\BibitemShut {NoStop}%
\bibitem [{\citenamefont {Moudgalya}\ \emph {et~al.}(2021)\citenamefont
  {Moudgalya}, \citenamefont {Bernevig},\ and\ \citenamefont
  {Regnault}}]{moudgalya2021quantum}%
  \BibitemOpen
  \bibfield  {author} {\bibinfo {author} {\bibfnamefont {S.}~\bibnamefont
  {Moudgalya}}, \bibinfo {author} {\bibfnamefont {B.~A.}\ \bibnamefont
  {Bernevig}},\ and\ \bibinfo {author} {\bibfnamefont {N.}~\bibnamefont
  {Regnault}},\ }\bibfield  {journal} {\bibinfo  {journal} {arXiv preprint
  arXiv:2109.00548}\ }\href
  {https://doi.org/https://doi.org/10.1088/1361-6633/ac73a0}
  {https://doi.org/10.1088/1361-6633/ac73a0} (\bibinfo {year}
  {2021})\BibitemShut {NoStop}%
\bibitem [{\citenamefont {Choi}\ \emph {et~al.}(2019)\citenamefont {Choi},
  \citenamefont {Turner}, \citenamefont {Pichler}, \citenamefont {Ho},
  \citenamefont {Michailidis}, \citenamefont {Papi\ifmmode~\acute{c}\else
  \'{c}\fi{}}, \citenamefont {Serbyn}, \citenamefont {Lukin},\ and\
  \citenamefont {Abanin}}]{Choi2019emergent}%
  \BibitemOpen
  \bibfield  {author} {\bibinfo {author} {\bibfnamefont {S.}~\bibnamefont
  {Choi}}, \bibinfo {author} {\bibfnamefont {C.~J.}\ \bibnamefont {Turner}},
  \bibinfo {author} {\bibfnamefont {H.}~\bibnamefont {Pichler}}, \bibinfo
  {author} {\bibfnamefont {W.~W.}\ \bibnamefont {Ho}}, \bibinfo {author}
  {\bibfnamefont {A.~A.}\ \bibnamefont {Michailidis}}, \bibinfo {author}
  {\bibfnamefont {Z.}~\bibnamefont {Papi\ifmmode~\acute{c}\else \'{c}\fi{}}},
  \bibinfo {author} {\bibfnamefont {M.}~\bibnamefont {Serbyn}}, \bibinfo
  {author} {\bibfnamefont {M.~D.}\ \bibnamefont {Lukin}},\ and\ \bibinfo
  {author} {\bibfnamefont {D.~A.}\ \bibnamefont {Abanin}},\ }\href
  {https://doi.org/10.1103/PhysRevLett.122.220603} {\bibfield  {journal}
  {\bibinfo  {journal} {Phys. Rev. Lett.}\ }\textbf {\bibinfo {volume} {122}},\
  \bibinfo {pages} {220603} (\bibinfo {year} {2019})}\BibitemShut {NoStop}%
\bibitem [{\citenamefont {Chandran}\ \emph {et~al.}(2023)\citenamefont
  {Chandran}, \citenamefont {Iadecola}, \citenamefont {Khemani},\ and\
  \citenamefont {Moessner}}]{Chandran2023qmbs}%
  \BibitemOpen
  \bibfield  {author} {\bibinfo {author} {\bibfnamefont {A.}~\bibnamefont
  {Chandran}}, \bibinfo {author} {\bibfnamefont {T.}~\bibnamefont {Iadecola}},
  \bibinfo {author} {\bibfnamefont {V.}~\bibnamefont {Khemani}},\ and\ \bibinfo
  {author} {\bibfnamefont {R.}~\bibnamefont {Moessner}},\ }\href
  {https://doi.org/10.1146/annurev-conmatphys-031620-101617} {\bibfield
  {journal} {\bibinfo  {journal} {Annual Review of Condensed Matter Physics}\
  }\textbf {\bibinfo {volume} {14}},\ \bibinfo {pages} {null} (\bibinfo {year}
  {2023})},\ \Eprint
  {https://arxiv.org/abs/https://doi.org/10.1146/annurev-conmatphys-031620-101617}
  {https://doi.org/10.1146/annurev-conmatphys-031620-101617} \BibitemShut
  {NoStop}%
\bibitem [{\citenamefont {Desaules}\ \emph {et~al.}(2021)\citenamefont
  {Desaules}, \citenamefont {Hudomal}, \citenamefont {Turner},\ and\
  \citenamefont {Papi\ifmmode~\acute{c}\else
  \'{c}\fi{}}}]{Desaules2021proposal}%
  \BibitemOpen
  \bibfield  {author} {\bibinfo {author} {\bibfnamefont {J.-Y.}\ \bibnamefont
  {Desaules}}, \bibinfo {author} {\bibfnamefont {A.}~\bibnamefont {Hudomal}},
  \bibinfo {author} {\bibfnamefont {C.~J.}\ \bibnamefont {Turner}},\ and\
  \bibinfo {author} {\bibfnamefont {Z.}~\bibnamefont
  {Papi\ifmmode~\acute{c}\else \'{c}\fi{}}},\ }\href
  {https://doi.org/10.1103/PhysRevLett.126.210601} {\bibfield  {journal}
  {\bibinfo  {journal} {Phys. Rev. Lett.}\ }\textbf {\bibinfo {volume} {126}},\
  \bibinfo {pages} {210601} (\bibinfo {year} {2021})}\BibitemShut {NoStop}%
\bibitem [{\citenamefont {Khemani}\ \emph {et~al.}(2020)\citenamefont
  {Khemani}, \citenamefont {Hermele},\ and\ \citenamefont
  {Nandkishore}}]{Khemani2020localization}%
  \BibitemOpen
  \bibfield  {author} {\bibinfo {author} {\bibfnamefont {V.}~\bibnamefont
  {Khemani}}, \bibinfo {author} {\bibfnamefont {M.}~\bibnamefont {Hermele}},\
  and\ \bibinfo {author} {\bibfnamefont {R.}~\bibnamefont {Nandkishore}},\
  }\href {https://doi.org/10.1103/PhysRevB.101.174204} {\bibfield  {journal}
  {\bibinfo  {journal} {Phys. Rev. B}\ }\textbf {\bibinfo {volume} {101}},\
  \bibinfo {pages} {174204} (\bibinfo {year} {2020})}\BibitemShut {NoStop}%
\bibitem [{\citenamefont {Wildeboer}\ \emph {et~al.}(2022)\citenamefont
  {Wildeboer}, \citenamefont {Langlett}, \citenamefont {Yang}, \citenamefont
  {Gorshkov}, \citenamefont {Iadecola},\ and\ \citenamefont
  {Xu}}]{Wildeboer2022qmbs}%
  \BibitemOpen
  \bibfield  {author} {\bibinfo {author} {\bibfnamefont {J.}~\bibnamefont
  {Wildeboer}}, \bibinfo {author} {\bibfnamefont {C.~M.}\ \bibnamefont
  {Langlett}}, \bibinfo {author} {\bibfnamefont {Z.-C.}\ \bibnamefont {Yang}},
  \bibinfo {author} {\bibfnamefont {A.~V.}\ \bibnamefont {Gorshkov}}, \bibinfo
  {author} {\bibfnamefont {T.}~\bibnamefont {Iadecola}},\ and\ \bibinfo
  {author} {\bibfnamefont {S.}~\bibnamefont {Xu}},\ }\href
  {https://doi.org/10.1103/PhysRevB.106.205142} {\bibfield  {journal} {\bibinfo
   {journal} {Phys. Rev. B}\ }\textbf {\bibinfo {volume} {106}},\ \bibinfo
  {pages} {205142} (\bibinfo {year} {2022})}\BibitemShut {NoStop}%
\bibitem [{\citenamefont {Zhang}\ \emph {et~al.}(2023)\citenamefont {Zhang},
  \citenamefont {Dong}, \citenamefont {Gao}, \citenamefont {Zhao},
  \citenamefont {Hao}, \citenamefont {Desaules}, \citenamefont {Guo},
  \citenamefont {Chen}, \citenamefont {Deng}, \citenamefont {Liu},
  \citenamefont {Ren}, \citenamefont {Yao}, \citenamefont {Zhang},
  \citenamefont {Xu}, \citenamefont {Wang}, \citenamefont {Jin}, \citenamefont
  {Zhu}, \citenamefont {Zhang}, \citenamefont {Li}, \citenamefont {Song},
  \citenamefont {Wang}, \citenamefont {Liu}, \citenamefont {Papi{\'{c}}},
  \citenamefont {Ying}, \citenamefont {Wang},\ and\ \citenamefont
  {Lai}}]{Zhang2023}%
  \BibitemOpen
  \bibfield  {author} {\bibinfo {author} {\bibfnamefont {P.}~\bibnamefont
  {Zhang}}, \bibinfo {author} {\bibfnamefont {H.}~\bibnamefont {Dong}},
  \bibinfo {author} {\bibfnamefont {Y.}~\bibnamefont {Gao}}, \bibinfo {author}
  {\bibfnamefont {L.}~\bibnamefont {Zhao}}, \bibinfo {author} {\bibfnamefont
  {J.}~\bibnamefont {Hao}}, \bibinfo {author} {\bibfnamefont {J.-Y.}\
  \bibnamefont {Desaules}}, \bibinfo {author} {\bibfnamefont {Q.}~\bibnamefont
  {Guo}}, \bibinfo {author} {\bibfnamefont {J.}~\bibnamefont {Chen}}, \bibinfo
  {author} {\bibfnamefont {J.}~\bibnamefont {Deng}}, \bibinfo {author}
  {\bibfnamefont {B.}~\bibnamefont {Liu}}, \bibinfo {author} {\bibfnamefont
  {W.}~\bibnamefont {Ren}}, \bibinfo {author} {\bibfnamefont {Y.}~\bibnamefont
  {Yao}}, \bibinfo {author} {\bibfnamefont {X.}~\bibnamefont {Zhang}}, \bibinfo
  {author} {\bibfnamefont {S.}~\bibnamefont {Xu}}, \bibinfo {author}
  {\bibfnamefont {K.}~\bibnamefont {Wang}}, \bibinfo {author} {\bibfnamefont
  {F.}~\bibnamefont {Jin}}, \bibinfo {author} {\bibfnamefont {X.}~\bibnamefont
  {Zhu}}, \bibinfo {author} {\bibfnamefont {B.}~\bibnamefont {Zhang}}, \bibinfo
  {author} {\bibfnamefont {H.}~\bibnamefont {Li}}, \bibinfo {author}
  {\bibfnamefont {C.}~\bibnamefont {Song}}, \bibinfo {author} {\bibfnamefont
  {Z.}~\bibnamefont {Wang}}, \bibinfo {author} {\bibfnamefont {F.}~\bibnamefont
  {Liu}}, \bibinfo {author} {\bibfnamefont {Z.}~\bibnamefont {Papi{\'{c}}}},
  \bibinfo {author} {\bibfnamefont {L.}~\bibnamefont {Ying}}, \bibinfo {author}
  {\bibfnamefont {H.}~\bibnamefont {Wang}},\ and\ \bibinfo {author}
  {\bibfnamefont {Y.-C.}\ \bibnamefont {Lai}},\ }\href
  {https://doi.org/10.1038/s41567-022-01784-9} {\bibfield  {journal} {\bibinfo
  {journal} {Nature Physics}\ }\textbf {\bibinfo {volume} {19}},\ \bibinfo
  {pages} {120} (\bibinfo {year} {2023})}\BibitemShut {NoStop}%
\bibitem [{\citenamefont {Moudgalya}\ \emph
  {et~al.}(2018{\natexlab{a}})\citenamefont {Moudgalya}, \citenamefont
  {Rachel}, \citenamefont {Bernevig},\ and\ \citenamefont
  {Regnault}}]{Moudgalya2018exact}%
  \BibitemOpen
  \bibfield  {author} {\bibinfo {author} {\bibfnamefont {S.}~\bibnamefont
  {Moudgalya}}, \bibinfo {author} {\bibfnamefont {S.}~\bibnamefont {Rachel}},
  \bibinfo {author} {\bibfnamefont {B.~A.}\ \bibnamefont {Bernevig}},\ and\
  \bibinfo {author} {\bibfnamefont {N.}~\bibnamefont {Regnault}},\ }\href
  {https://doi.org/10.1103/PhysRevB.98.235155} {\bibfield  {journal} {\bibinfo
  {journal} {Phys. Rev. B}\ }\textbf {\bibinfo {volume} {98}},\ \bibinfo
  {pages} {235155} (\bibinfo {year} {2018}{\natexlab{a}})}\BibitemShut
  {NoStop}%
\bibitem [{\citenamefont {Moudgalya}\ \emph
  {et~al.}(2018{\natexlab{b}})\citenamefont {Moudgalya}, \citenamefont
  {Regnault},\ and\ \citenamefont {Bernevig}}]{Moudgalya2018AKLT}%
  \BibitemOpen
  \bibfield  {author} {\bibinfo {author} {\bibfnamefont {S.}~\bibnamefont
  {Moudgalya}}, \bibinfo {author} {\bibfnamefont {N.}~\bibnamefont
  {Regnault}},\ and\ \bibinfo {author} {\bibfnamefont {B.~A.}\ \bibnamefont
  {Bernevig}},\ }\href {https://doi.org/10.1103/PhysRevB.98.235156} {\bibfield
  {journal} {\bibinfo  {journal} {Phys. Rev. B}\ }\textbf {\bibinfo {volume}
  {98}},\ \bibinfo {pages} {235156} (\bibinfo {year}
  {2018}{\natexlab{b}})}\BibitemShut {NoStop}%
\bibitem [{\citenamefont {Mark}\ \emph {et~al.}(2020)\citenamefont {Mark},
  \citenamefont {Lin},\ and\ \citenamefont {Motrunich}}]{Mark2020AKLT}%
  \BibitemOpen
  \bibfield  {author} {\bibinfo {author} {\bibfnamefont {D.~K.}\ \bibnamefont
  {Mark}}, \bibinfo {author} {\bibfnamefont {C.-J.}\ \bibnamefont {Lin}},\ and\
  \bibinfo {author} {\bibfnamefont {O.~I.}\ \bibnamefont {Motrunich}},\ }\href
  {https://doi.org/10.1103/PhysRevB.101.195131} {\bibfield  {journal} {\bibinfo
   {journal} {Phys. Rev. B}\ }\textbf {\bibinfo {volume} {101}},\ \bibinfo
  {pages} {195131} (\bibinfo {year} {2020})}\BibitemShut {NoStop}%
\bibitem [{\citenamefont {Chattopadhyay}\ \emph {et~al.}(2020)\citenamefont
  {Chattopadhyay}, \citenamefont {Pichler}, \citenamefont {Lukin},\ and\
  \citenamefont {Ho}}]{Chattopadhyay2020qmbs}%
  \BibitemOpen
  \bibfield  {author} {\bibinfo {author} {\bibfnamefont {S.}~\bibnamefont
  {Chattopadhyay}}, \bibinfo {author} {\bibfnamefont {H.}~\bibnamefont
  {Pichler}}, \bibinfo {author} {\bibfnamefont {M.~D.}\ \bibnamefont {Lukin}},\
  and\ \bibinfo {author} {\bibfnamefont {W.~W.}\ \bibnamefont {Ho}},\ }\href
  {https://doi.org/10.1103/PhysRevB.101.174308} {\bibfield  {journal} {\bibinfo
   {journal} {Phys. Rev. B}\ }\textbf {\bibinfo {volume} {101}},\ \bibinfo
  {pages} {174308} (\bibinfo {year} {2020})}\BibitemShut {NoStop}%
\bibitem [{\citenamefont {Schecter}\ and\ \citenamefont
  {Iadecola}(2019)}]{Schecter2019weak}%
  \BibitemOpen
  \bibfield  {author} {\bibinfo {author} {\bibfnamefont {M.}~\bibnamefont
  {Schecter}}\ and\ \bibinfo {author} {\bibfnamefont {T.}~\bibnamefont
  {Iadecola}},\ }\href {https://doi.org/10.1103/PhysRevLett.123.147201}
  {\bibfield  {journal} {\bibinfo  {journal} {Phys. Rev. Lett.}\ }\textbf
  {\bibinfo {volume} {123}},\ \bibinfo {pages} {147201} (\bibinfo {year}
  {2019})}\BibitemShut {NoStop}%
\bibitem [{\citenamefont {Moudgalya}\ \emph
  {et~al.}(2020{\natexlab{a}})\citenamefont {Moudgalya}, \citenamefont
  {Regnault},\ and\ \citenamefont {Bernevig}}]{Moudgalya2020hubbard}%
  \BibitemOpen
  \bibfield  {author} {\bibinfo {author} {\bibfnamefont {S.}~\bibnamefont
  {Moudgalya}}, \bibinfo {author} {\bibfnamefont {N.}~\bibnamefont
  {Regnault}},\ and\ \bibinfo {author} {\bibfnamefont {B.~A.}\ \bibnamefont
  {Bernevig}},\ }\href {https://doi.org/10.1103/PhysRevB.102.085140} {\bibfield
   {journal} {\bibinfo  {journal} {Phys. Rev. B}\ }\textbf {\bibinfo {volume}
  {102}},\ \bibinfo {pages} {085140} (\bibinfo {year}
  {2020}{\natexlab{a}})}\BibitemShut {NoStop}%
\bibitem [{\citenamefont {Mark}\ and\ \citenamefont
  {Motrunich}(2020)}]{Mark2020hubbard}%
  \BibitemOpen
  \bibfield  {author} {\bibinfo {author} {\bibfnamefont {D.~K.}\ \bibnamefont
  {Mark}}\ and\ \bibinfo {author} {\bibfnamefont {O.~I.}\ \bibnamefont
  {Motrunich}},\ }\href {https://doi.org/10.1103/PhysRevB.102.075132}
  {\bibfield  {journal} {\bibinfo  {journal} {Phys. Rev. B}\ }\textbf {\bibinfo
  {volume} {102}},\ \bibinfo {pages} {075132} (\bibinfo {year}
  {2020})}\BibitemShut {NoStop}%
\bibitem [{\citenamefont {Bull}\ \emph {et~al.}(2019)\citenamefont {Bull},
  \citenamefont {Martin},\ and\ \citenamefont {Papi\ifmmode~\acute{c}\else
  \'{c}\fi{}}}]{Bull2019systematic}%
  \BibitemOpen
  \bibfield  {author} {\bibinfo {author} {\bibfnamefont {K.}~\bibnamefont
  {Bull}}, \bibinfo {author} {\bibfnamefont {I.}~\bibnamefont {Martin}},\ and\
  \bibinfo {author} {\bibfnamefont {Z.}~\bibnamefont
  {Papi\ifmmode~\acute{c}\else \'{c}\fi{}}},\ }\href
  {https://doi.org/10.1103/PhysRevLett.123.030601} {\bibfield  {journal}
  {\bibinfo  {journal} {Phys. Rev. Lett.}\ }\textbf {\bibinfo {volume} {123}},\
  \bibinfo {pages} {030601} (\bibinfo {year} {2019})}\BibitemShut {NoStop}%
\bibitem [{\citenamefont {Moudgalya}\ \emph
  {et~al.}(2020{\natexlab{b}})\citenamefont {Moudgalya}, \citenamefont
  {Bernevig},\ and\ \citenamefont {Regnault}}]{Moudgalya2020thintorus}%
  \BibitemOpen
  \bibfield  {author} {\bibinfo {author} {\bibfnamefont {S.}~\bibnamefont
  {Moudgalya}}, \bibinfo {author} {\bibfnamefont {B.~A.}\ \bibnamefont
  {Bernevig}},\ and\ \bibinfo {author} {\bibfnamefont {N.}~\bibnamefont
  {Regnault}},\ }\href {https://doi.org/10.1103/PhysRevB.102.195150} {\bibfield
   {journal} {\bibinfo  {journal} {Phys. Rev. B}\ }\textbf {\bibinfo {volume}
  {102}},\ \bibinfo {pages} {195150} (\bibinfo {year}
  {2020}{\natexlab{b}})}\BibitemShut {NoStop}%
\bibitem [{\citenamefont {Kitaev}(2006)}]{Kitaev2006anyons}%
  \BibitemOpen
  \bibfield  {author} {\bibinfo {author} {\bibfnamefont {A.}~\bibnamefont
  {Kitaev}},\ }\href
  {https://doi.org/https://doi.org/10.1016/j.aop.2005.10.005} {\bibfield
  {journal} {\bibinfo  {journal} {Annals of Physics}\ }\textbf {\bibinfo
  {volume} {321}},\ \bibinfo {pages} {2} (\bibinfo {year} {2006})},\ \bibinfo
  {note} {january Special Issue}\BibitemShut {NoStop}%
\bibitem [{\citenamefont {Sen}\ \emph {et~al.}(2010)\citenamefont {Sen},
  \citenamefont {Shankar}, \citenamefont {Dhar},\ and\ \citenamefont
  {Ramola}}]{Sen2010kitaev}%
  \BibitemOpen
  \bibfield  {author} {\bibinfo {author} {\bibfnamefont {D.}~\bibnamefont
  {Sen}}, \bibinfo {author} {\bibfnamefont {R.}~\bibnamefont {Shankar}},
  \bibinfo {author} {\bibfnamefont {D.}~\bibnamefont {Dhar}},\ and\ \bibinfo
  {author} {\bibfnamefont {K.}~\bibnamefont {Ramola}},\ }\href
  {https://doi.org/10.1103/PhysRevB.82.195435} {\bibfield  {journal} {\bibinfo
  {journal} {Phys. Rev. B}\ }\textbf {\bibinfo {volume} {82}},\ \bibinfo
  {pages} {195435} (\bibinfo {year} {2010})}\BibitemShut {NoStop}%
\bibitem [{\citenamefont {You}\ \emph {et~al.}(2022)\citenamefont {You},
  \citenamefont {Zhao}, \citenamefont {Ren}, \citenamefont {Sun}, \citenamefont
  {Li},\ and\ \citenamefont {Ole\ifmmode~\acute{s}\else
  \'{s}\fi{}}}]{You2022qmbs}%
  \BibitemOpen
  \bibfield  {author} {\bibinfo {author} {\bibfnamefont {W.-L.}\ \bibnamefont
  {You}}, \bibinfo {author} {\bibfnamefont {Z.}~\bibnamefont {Zhao}}, \bibinfo
  {author} {\bibfnamefont {J.}~\bibnamefont {Ren}}, \bibinfo {author}
  {\bibfnamefont {G.}~\bibnamefont {Sun}}, \bibinfo {author} {\bibfnamefont
  {L.}~\bibnamefont {Li}},\ and\ \bibinfo {author} {\bibfnamefont {A.~M.}\
  \bibnamefont {Ole\ifmmode~\acute{s}\else \'{s}\fi{}}},\ }\href
  {https://doi.org/10.1103/PhysRevResearch.4.013103} {\bibfield  {journal}
  {\bibinfo  {journal} {Phys. Rev. Res.}\ }\textbf {\bibinfo {volume} {4}},\
  \bibinfo {pages} {013103} (\bibinfo {year} {2022})}\BibitemShut {NoStop}%
\bibitem [{\citenamefont {Rozon}\ and\ \citenamefont
  {Agarwal}(2023)}]{Rozon22_broken_unitary}%
  \BibitemOpen
  \bibfield  {author} {\bibinfo {author} {\bibfnamefont {P.-G.}\ \bibnamefont
  {Rozon}}\ and\ \bibinfo {author} {\bibfnamefont {K.}~\bibnamefont
  {Agarwal}},\ }\href {https://doi.org/10.48550/ARXIV.2302.04885} {\bibinfo
  {title} {Broken unitary picture of dynamics in quantum many-body scars}}
  (\bibinfo {year} {2023})\BibitemShut {NoStop}%
\bibitem [{\citenamefont {Windt}\ and\ \citenamefont
  {Pichler}(2022)}]{Windt22_SqueezingQMBS}%
  \BibitemOpen
  \bibfield  {author} {\bibinfo {author} {\bibfnamefont {B.}~\bibnamefont
  {Windt}}\ and\ \bibinfo {author} {\bibfnamefont {H.}~\bibnamefont
  {Pichler}},\ }\href {https://doi.org/10.1103/PhysRevLett.128.090606}
  {\bibfield  {journal} {\bibinfo  {journal} {Phys. Rev. Lett.}\ }\textbf
  {\bibinfo {volume} {128}},\ \bibinfo {pages} {090606} (\bibinfo {year}
  {2022})}\BibitemShut {NoStop}%
\bibitem [{\citenamefont {Shiraishi}\ and\ \citenamefont
  {Mori}(2017)}]{Shiraishi17_projector_embedding}%
  \BibitemOpen
  \bibfield  {author} {\bibinfo {author} {\bibfnamefont {N.}~\bibnamefont
  {Shiraishi}}\ and\ \bibinfo {author} {\bibfnamefont {T.}~\bibnamefont
  {Mori}},\ }\href {https://doi.org/10.1103/PhysRevLett.119.030601} {\bibfield
  {journal} {\bibinfo  {journal} {Phys. Rev. Lett.}\ }\textbf {\bibinfo
  {volume} {119}},\ \bibinfo {pages} {030601} (\bibinfo {year}
  {2017})}\BibitemShut {NoStop}%
\bibitem [{\citenamefont {Moudgalya}\ and\ \citenamefont
  {Motrunich}(2022)}]{Moudgalya22_commutant_algebras}%
  \BibitemOpen
  \bibfield  {author} {\bibinfo {author} {\bibfnamefont {S.}~\bibnamefont
  {Moudgalya}}\ and\ \bibinfo {author} {\bibfnamefont {O.~I.}\ \bibnamefont
  {Motrunich}},\ }\href {https://doi.org/10.1103/PhysRevX.12.011050} {\bibfield
   {journal} {\bibinfo  {journal} {Phys. Rev. X}\ }\textbf {\bibinfo {volume}
  {12}},\ \bibinfo {pages} {011050} (\bibinfo {year} {2022})}\BibitemShut
  {NoStop}%
\end{thebibliography}%
\bibliographystyle{apsrev4-2}
\end{document}